\documentclass[]{pasj01}
\draft

\Received{}
\Accepted{}
 
\begin{document} 

\title{Formation of High-Mass stars in an isolated environment in the Large Magellanic Cloud
}

\author{Ryohei \textsc{Harada}\altaffilmark{1}%
}
\altaffiltext{1}{Department of Physical Science, Osaka Prefecture University, 1-1 Gakuen-cho, Naka-ku, Sakai, Osaka 599-8531, Japan}
\email{s-r.harada@p.s.osakafu-u.ac.jp}

\author{Toshikazu \textsc{Onishi}\altaffilmark{1}}

\author{Kazuki \textsc{Tokuda}\altaffilmark{1,2}}
\altaffiltext{2}{Chile Observatory, National Astronomical Observatory of Japan, National Institutes of Natural Science, 2-21-1 Osawa, Mitaka, Tokyo
181-8588, Japan}

\author{Sarolta \textsc{Zahorecz}\altaffilmark{1,2}}

\author{Annie \textsc{Hughes}\altaffilmark{3,4}}
\altaffiltext{3}{CNRS, IRAP, 9 Av. du Colonel Roche, BP 44346, F-31028 Toulouse cedex 4, France}
\altaffiltext{4}{Universit\'{e} de Toulouse, UPS-OMP, IRAP, F-31028 Toulouse cedex 4, France}

\author{Margaret \textsc{Meixner}\altaffilmark{5,6}}
\altaffiltext{5}{Space Telescope Science Institute, 3700 San Martin Drive, Baltimore, MD 21218, USA}
\altaffiltext{6}{Department of Physics \& Astronomy, Johns Hopkins University, 3400 N. Charles Street, Baltimore, MD 21218, USA}

\author{Marta \textsc{Sewi{\l}o}\altaffilmark{7,8}}
\altaffiltext{7}{CRESST II and Exoplanets and Stellar Astrophysics Laboratory, NASA Goddard Space Flight Center, Greenbelt, MD 20771, USA}
\altaffiltext{8}{Department of Astronomy, University of Maryland, College Park, MD 20742, USA}

\author{Remy \textsc{Indebetouw}\altaffilmark{9,10}}
\altaffiltext{9}{Department of Astronomy, University of Virginia, P.O. Box 400325, Charlottesville, VA 22904, USA}
\altaffiltext{10}{National Radio Astronomy Observatory, 520 Edgemont Road, Charlottesville, VA 22903, USA}

\author{Omnarayani \textsc{Nayak}\altaffilmark{6}}

\author{Yasuo \textsc{Fukui}\altaffilmark{11}}
\altaffiltext{11}{Department of Physics, Nagoya University, Chikusa-ku, Nagoya 464-8602, Japan}

\author{Kengo \textsc{Tachihara}\altaffilmark{11}}

\author{Kisetstu \textsc{Tsuge}\altaffilmark{11}}

\author{Akiko \textsc{Kawamura}\altaffilmark{2}}

\author{Kazuya \textsc{Saigo}\altaffilmark{2}}

\author{Tony \textsc{Wong}\altaffilmark{12}}
\altaffiltext{12}{Astronomy Department, University of Illinois, Urbana, IL 61801, USA}

\author{Jean-Philippe \textsc{Bernard}\altaffilmark{3,4}}

\author{Ian \textsc{W. Stephens}\altaffilmark{13}}
\altaffiltext{13}{Harvard-Smithsonian Center for Astrophysics, 60 Garden Street, Cambridge, MA, USA}



\maketitle

\begin{abstract}
The aim of this study is to characterize the distribution and basic properties of the natal gas associated with high-mass young stellar objects (YSOs) in isolated environments in the Large Magellanic Cloud (LMC).  High-mass stars usually form in Giant Molecular Clouds (GMCs) as part of a young stellar cluster, but some OB stars are observed far from GMCs.  By examining the spatial coincidence between the high-mass YSOs and $^{12}$CO ($J$ = 1--0) emission detected by NANTEN and Mopra observations, we selected ten high-mass YSOs that are located away from any of the NANTEN clouds but are detected by the Mopra pointed observations.  The ALMA observations revealed that a compact molecular cloud whose mass is a few thousand solar masses or smaller is associated with  the high-mass YSOs, which indicates that these compact clouds are the sites of high-mass star formation.  The high-density and high-temperature throughout the clouds are explained by the severe photodissociation of CO due to the lower metallicity than in the Galaxy.  The star formation efficiency ranges from several to as high as $\sim$$\;$40\%, indicating efficient star formation in these environments.  The enhanced turbulence may be a cause of the efficient star formation therein, as judged from the gas velocity information and the association with the lower density gas.
\end{abstract}

\section{Introduction}
High-mass stars strongly influence physically and chemically the interstellar matter and thus galactic evolution.  They emit a significant amount of ultraviolet emission, ionizing and heating the ambient gas.  They also generate a stellar wind, and it compresses the surrounding interstellar medium.  Furthermore, a supernova explosion occurs at the end of their evolution, releasing massive energy into interstellar space.  The interstellar matter is strongly influenced by the released energy from the high-mass star.  It is, therefore, extremely important to investigate the formation mechanism of high-mass stars to study galaxy evolution.
Our current theory of star formation proposes that: 1) most stars are formed in giant molecular clouds (GMCs) as part of a cluster or group of stars, 2) the evolutionary timescale of high-mass young stellar objects (YSOs) is short compared to the disruption timescale of their parent molecular cloud, and 3) star formation is inefficient (i.e., a low fraction of the molecular gas mass is ultimately converted into stars) (c.f., \cite{Zinnecker07} and the references therein).  Assuming a normal stellar initial mass function, the birth of a single high-mass star should therefore always be coincident with the formation of many low-mass YSOs.  The formation mechanism of a high-mass star, however, is not straightforward; high-mass stars are considered to be formed by the collapse of a highly turbulent core (core accretion models; e.g., \cite{McKee02, McKee03}) or by Bondi-Hoyle accretion onto small protostars (competitive accretion model; e.g., \cite{Zinnecker82}, \cite{Bonnell97}).  Molecular observations toward the natal clouds of high-mass YSOs are of vital importance to resolve the issue although high-mass star-forming regions are usually embedded in a complex environment in the plane of the galaxy.\\
\ \ Although most OB stars are believed to be formed as associations/clusters in GMCs, there exists a class of OB stars that are distant from such complexes.  Many are considered to be runaway OB stars that are ejected from their birthplace; however, some cannot be assigned to a probable progenitor cloud and the origin of these $``$isolated$"$ OB stars located far from molecular cloud complexes remains unknown (e.g., \cite{de Wit04, de Wit05}; \cite{Zinnecker07} and the references therein). \\
\ \ The majority of Milky Way YSOs are located along the Galactic plane, and therefore distance ambiguity and contamination from unrelated emission sources along the same line-of-sight complicates the study of Galactic high-mass YSOs.  With the high resolution and sensitivity of ALMA, we can now study CO gas in nearby galaxies in great detail.  The Large Magellanic Cloud (LMC) is located close to us ($\sim$$\;$50$\>$kpc; \cite{Schaefer08}; \cite{de Grijs14}) at high Galactic latitude ($b$ $\sim$$\;$$-\timeform{30D}$) with face-on orientation ($i$ $\sim$$\;$$\timeform{38D}$; \cite{Balbinot15}).  Therefore, toward the LMC, we have the clearest view of the distribution of molecular clouds and young stars in any galaxy, including our own, making the LMC one of the best places to investigate the origin of high-mass YSOs.\\
\ \ The CO clouds in the LMC have been studied extensively as reviewed by \citet{Fukui10} (see references therein).  Spatially-resolved observations of the GMCs at 40$\>$pc resolution in the whole LMC disk were initiated with NANTEN by \citet{Fukui99}, and followed up at higher sensitivity by \citet{Fukui08} and \citet{Kawamura09}.  Subsequently, higher resolution CO observations with Mopra telescope by \citet{Wong11} revealed the CO distribution at 11$\>$pc resolution toward the NANTEN GMCs and their physical properties were studied in detail (\cite{Hughes10}).  These preceding works focused on massive and active OB star-forming regions with clusters, whereas it was not explored how star formation is taking place at smaller scales, involving only a single OB star.  It is possible that such small-scale star formation is important in galactic evolution if they are numerous in the whole disk.  It is also important to learn the implications of single OB star formation in theories of high mass star formation, which assume massive aggregates as the precursor of O star clusters (e.g., see for a review \cite{Tan14}; also \cite{Ascenso18}).\\ 
\ \ In this paper, we present the results of ALMA observations of compact molecular clouds associated with $``$isolated$"$ high-mass YSOs in the LMC that are located away from any GMCs to show the physical properties of the natal clouds.  Since high-mass star formation involves many diverse processes, the study of their formation can be facilitated by characterizing the molecular gas of more isolated clouds. This study specifically targets high-mass YSOs that are encompassed by relatively isolated clouds. We characterize these clouds in detail based on ALMA and ancillary observations, and make comparisons to known Galactic star-forming regions.

\newpage

\section{Source selection and observations} \label{sec:observations}
\begin{figure}[ht!]
 \begin{center}
  \includegraphics[width=120mm]{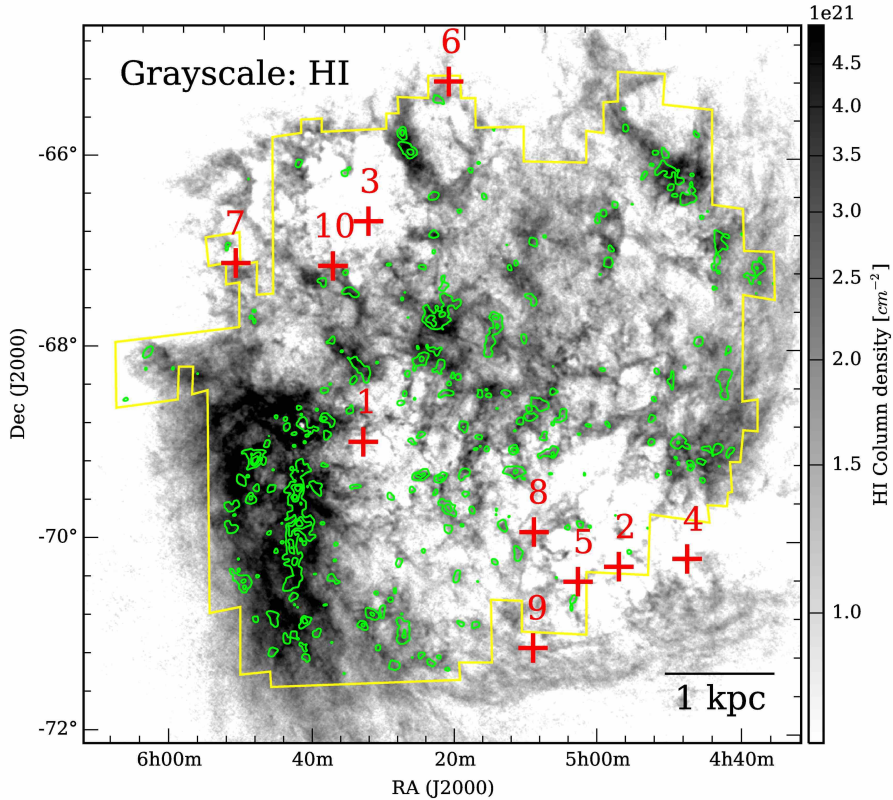}
 \end{center}
\caption{Positions of the isolated high-mass YSOs in the Large Magellanic Cloud observed by our ALMA observations (red crosses with numbers) overlaid on the H$\;${\sc i} column density map \citep{Kim03}.  Green contours show CO ($J$ = 1--0) integrated intensities with the NANTEN telescope \citep{Fukui08}.  The yellow lines show the coverage of the NANTEN survey. 
\label{fig:nanten_co}}
\end{figure}

As part of the Surveying the Agents of a Galaxy's Evolution (SAGE) {\it Spitzer} Legacy program \citep{Meixner06}, $\sim$$\;$1,800 unique YSO candidates were cataloged in the LMC (\cite{Whitney08}; see also \cite{Gruendl09}).  The YSOs are selected by their excess of infrared emission, indicative of an early evolutionary stage when the YSOs are still surrounded by disks and/or infalling envelopes.  We identified high-mass YSOs in the LMC that appear to be isolated, i.e., not associated with CO emission by examining the spatial coincidence between the YSOs and $^{12}$CO ($J$ = 1--0) emission detected by the NANTEN survey (half-power beam width, HPBW $\sim\;$$\timeform{2'.6}$, \cite{Fukui08}) and MAGMA CO survey (\cite{Wong11}).  A further set of single-pointing CO observations with the Mopra Telescope (HPBW $\sim\;$$\timeform{30"}$; Proposal ID: M579, PI: T. Wong) targeted 76 $``$definite$"$ and $``$probable$"$ YSOs from \citet{Gruendl09} with [8.0]$\;<\;$8.0 mag that were outside the MAGMA-LMC survey coverage.  Single pointing observations toward the sources were carried out, with 80\% of sources detected in $^{12}$CO ($J$ = 1--0).  Follow-up Mopra CO mapping observations (Proposal ID: M2013B34, PI: T. Onishi) showed that CO emission was spatially extended toward some of the YSOs, although CO emission was only detected precisely at the YSO positions in some cases.  We note that a small fraction of the isolated YSOs with Mopra CO detections are located in regions with H$\;${\sc i} column densities of $\lesssim$ 10$^{21}$$\;$cm$^{-2}$, where equilibrium models of H$_2$ formation and dissociation (e.g., \cite{Krumholz09}) predict that there should be no molecular gas.\\
\ \ For this study, we targeted a typical sample, regarding their association with neutral gas tracers, of the isolated high-mass YSOs in the LMC.  All of the targets are separated by more than 200$\>$pc from NANTEN CO clouds to exclude potential $``$runaway$"$ sources, which are stars that have been ejected from their progenitor gas cloud at high speed.  This minimum spatial separation is obtained by assuming that these runaway YSOs have a maximum speed of 200$\;$km$\,$s$^{-1}$ (e.g., \cite{Perets12}) and a conservative age of 10$^6$$\;$yrs.  Included in the above samples are YSOs where the H$\;${\sc i} column density is less than 2$\;$$\times$$\;$10$^{21}$$\;$cm$^{-2}$ with an angular resolution of 1 arcmin \citep{Kim03}, which corresponds to the spatial resolution of 14.4$\>$pc.  Our final target sample (see table \ref{tab:sample}) is spatially distributed across the LMC gas disk (see figure \ref{fig:nanten_co}), and consists of:
\begin{enumerate}
\item[(a)] Three YSOs with CO emission that were unresolved by Mopra. The molecular clouds associated with these high mass YSOs must have sizes smaller than 7$\>$pc and masses less than a few thousand solar masses.

\item[(b)]  Seven YSOs where CO emission was detected by Mopra; CO mapping observations were not attempted for these sources, but the morphology of the 350$\;$$\mu$m emission suggests that the molecular clouds are likely to be spatially compact.  These CO clouds were not detected by NANTEN, which puts an upper limit on their mass of $\sim$$\;$10,000 solar masses \citep{Fukui08}.
\end{enumerate}
\ \ In order to determine the physical parameters of these YSOs, we used the YSO model grid of \authorcite{Robitaille06} (\yearcite{Robitaille06}, \yearcite{Robitaille07}) and the spectral energy distribution (SED) fitter for the available photometry data including {\it Spitzer} and {\it Herschel} fluxes (1.2--500$\;\mu$m).  Figure \ref{fig:nanten_co} shows the positions of the sources overlaid on the H$\;${\sc i} intensities \citep{Kim03}.  The basic properties are listed in table \ref{tab:sample}.  The luminosities range from 0.3 to 19$\;$$\times$$\;$10$^4$$\;$$\LO$, and the fitted masses from 10 to 17$\;$$M_{\odot}$, which means that they are B stars earlier than B2 (e.g., \cite{Hohle10}).  We note that these models do not include multiplicity even though most high-mass stars are multiples.
%
\begin{table}[ht!]
  \tbl{Characteristics of the observed isolated YSOs}{%
  \begin{tabular}{lcccccc}
      \hline
Target No. & \multicolumn{2}{c}{Position (J2000)} & Stellar Mass & Luminosity & H$\;${\sc i} column density & proposal ID \\
 & R.A. & Dec. & [$M_{\odot}$] & [$\times$$\;$10$^4$ $L_{\odot}$] & [$\times$$\;$10$^{21}$ cm$^{-2}$] & \\
 & & & (1) & (2) & (3) & (4) \\ 
      \hline
1 & $\timeform{5h31m27s.01}$ & -$\timeform{69D17'55".5}$ & 10$\pm$1 & 0.32$\pm$0.035 & 1.2 & M2013B34 \\
2 & $\timeform{4h59m23s.35}$ & -$\timeform{70D31'57".4}$ & 13 & 0.62 & 0.3 & M2013B34 \\
3 & $\timeform{5h30m07s.29}$ & -$\timeform{66D57'48".2}$ & 13$\pm$1 & 0.38$\pm$0.093 & 0.9 & M2013B34 \\
4 & $\timeform{4h50m54s.49}$ & -$\timeform{70D22'01".6}$ & 14$\pm$1 & 1.1$\pm$0.049 & 0.8 & M579 \\
5 & $\timeform{5h04m24s.82}$ & -$\timeform{70D43'43".7}$ & 17 & 19 & 1.5 & M579 \\
6--1 & $\timeform{5h21m33s.20}$ & -$\timeform{65D29'20".8}$ & 17$\pm$1 & 1.7$\pm$0.056 & 1.1 & M579 \\
6--2 & $\timeform{5h21m31s.11}$ & -$\timeform{65D29'14".8}$ & 13$\pm$2 & 1.1$\pm$0.35 & 0.9 & M579 \\
7 & $\timeform{5h44m49s.63}$ & -$\timeform{67D19'34".7}$ & 14$\pm$1 & 1.0$\pm$0.210 & 1.3 & M579 \\
8 & $\timeform{5h10m24s.09}$ & -$\timeform{70D14'06".5}$ &16$\pm$1 & 2.5$\pm$0.490 & 1.1 & M579 \\
9 & $\timeform{5h09m41s.94}$ & -$\timeform{71D27'42".2}$ & 15 & 1.2 & 1.7 & M579 \\
10 & $\timeform{5h34m10s.23}$ & -$\timeform{67D25'29".4}$ &15 & 1.1 & 1.6 & M579 \\
      \hline
    \end{tabular}}\label{tab:sample}
\begin{tabnote}
Col.(1)--(2): Masses and luminosities of the YSO.  No error for mass or luminosity means that there was only one model that fit the data points.  Col.(3): H$\;${\sc i} column densities at the YSO positions \citep{Kim03}.  Col.(4): Proposal ID of CO observation by the Mopra telescope.  See details in section \ref{sec:observations}. 
\end{tabnote}
\end{table}

\section{ALMA observations} \label{sec:alma_obs}
We carried out ALMA Cycle 2 Band 3 (86--116$\;$GHz) and Band 6 (211--275$\;$GHz) observations toward 10 isolated YSOs (Project 2013.1.00287.S, PI: Toshikazu Onishi).  The targeted molecular lines were $^{13}$CO ($J$ = 1--0), C$^{18}$O ($J$ = 1--0), CS ($J$ = 2--1), $^{12}$CO ($J$ = 2--1), $^{13}$CO ($J$ = 2--1) and C$^{18}$O($J$ = 2--1) with a frequency resolution of 61$\;$kHz, corresponding to velocity resolutions of $\sim\;$0.17$\;$km$\,$s$^{-1}$ for Band 3 and $\sim\;$0.08$\;$km$\,$s$^{-1}$ for Band 6, both with 1920 channels.  We also observed the continuum emission with the wider bandwidth of 1875$\;$MHz.  The H40$\alpha$ radio recombination line was included in the wider bandwidth observations with a frequency resolution of 976$\;$kHz ($\times$$\;$1920 channels).  The ALMA Band 6 observations were carried out in June 2014 and September 2015.  They used 34 antennas, and the projected baseline length of the 12$\>$m array ranges from 13$\>$m to 392$\>$m.  The ALMA Band 3 observations were carried out in December  2014.  They used 34 antennas and the projected baseline length of the 12$\>$m array ranges from 13$\>$m to 247$\>$m.  The projected baselines of the Atacama Compact Array (ACA; a.k.a. Morita Array) observation range from 8$\>$m to 36$\>$m.  For the Band 6 data, we combined the 12$\>$m array, 7$\>$m array, and TP array data to recover the extended emission except for Target 9, toward which we have no TP array observations due to the failed observations.  For Band 3, we didn't include the ACA observations because the maximum recoverable scale of $\sim$$\;$$\timeform{26"}$, corresponding to $\sim$$\;$6.2$\>$pc, is considered to be large enough to investigate the gas distribution of the compact clouds.  The ALMA beam sizes and sensitivities in the present observations are listed in table \ref{tab:beam_size}.
\begin{table}[ht!]
  \tbl{ALMA beam sizes and sensitivities}{%
  \begin{tabular}{lccccc}
      \hline
 & Band 3 &  \multicolumn{4}{c}{Band 6} \\
      \cline{3-6}
Parameters & 12$\>$m Array & 12$\>$m Array & 7$\>$m Array & TP & Combined \\
      \hline
Beam size [arcsec] & 3.2$\;$$\times$$\;$2.4  & 1.9$\;$$\times$$\;$1.1 & 7.0$\;$$\times$$\;$6.7  & 28.3 & 1.9$\;$$\times$$\;$1.1 \\
Sensitivity of line observation (rms)\footnotemark[$*$] [K] & $\sim$$\;$0.23 & $\sim$$\;$0.67 & $\sim$$\;$0.13 & $\sim$$\;$0.03 & $\sim$$\;$0.67\\
      \hline
    \end{tabular}}\label{tab:beam_size}
\begin{tabnote}
\footnotemark[$*$] The sensitivities are measured at a common velocity resolution of 0.2 km$\,$s$^{-1}$.  \\ 
\end{tabnote}
\end{table}

\section{Results} \label{sec:Results}
\subsection{Compact clouds associated with isolated YSOs } \label{subsec:compact_cloud}
We show the ALMA images toward the targets both in Band 3 and Band 6 in figure \ref{fig:ii_target01-03}.  CO and $^{13}$CO emission are detected from all the targets, and compact molecular clouds are associated with all the candidate isolated YSOs.  This confirms that they are indeed YSOs as they still have their natal clouds around them.  The figures show that the molecular distribution of $^{12}$CO ($J$ = 2--1) and $^{13}$CO ($J$ = 1--0) toward Target 1, 2, and 3 is compact having a size of $\sim$$\;$1$\>$pc, which is consistent with the past Mopra observations, and each YSO is located at the peak of the molecular emission.  The 8$\;\mu$m emission by {\it Spitzer} also shows a good correspondence with the molecular distribution.  We detected CS ($J$ = 2--1) emission toward Target 2, 5, 6, 8, and 10, indicating the association of dense gas toward these YSOs.  Two observation fields, Target 5 and 6, include multiple YSO candidates.  The observation field of Target 6 includes 3 YSOs, two of which are associated with molecular gas centered at the position of each YSO (6--1 and 6--2 in figure \ref{fig:ii_target01-03}).  The other source of Target 6 and one of the two sources of Target 5 are clearly not associated with molecular emission, indicating that they are more evolved YSOs or not YSOs.  Mopra single pointing observations toward Target 4--10 were not able to reveal the size of the CO clouds.  The present ALMA observations show a very compact cloud associated with Target 10, and some extended emission with a size of $\sim$$\;$5$\>$pc is associated with Target 4, 5, and 8.  For Target 4 and 7, the YSOs are located at the holes of the molecular distribution and H$\alpha$ emission with a size of $\sim$$\;$1$\>$pc is associated with these sources (see figure \ref{fig:continuum}).  This fact suggests that significant dissipation of the associated gas has started toward these two sources.  We detected Band 3 100$\;$GHz continuum emission as shown in figure \ref{fig:continuum} toward Target 5 and 6, and detected the H40$\alpha$ radio recombination line toward Target 6.

\begin{figure}[ht!]
 \begin{center}
  \includegraphics[width=140mm]{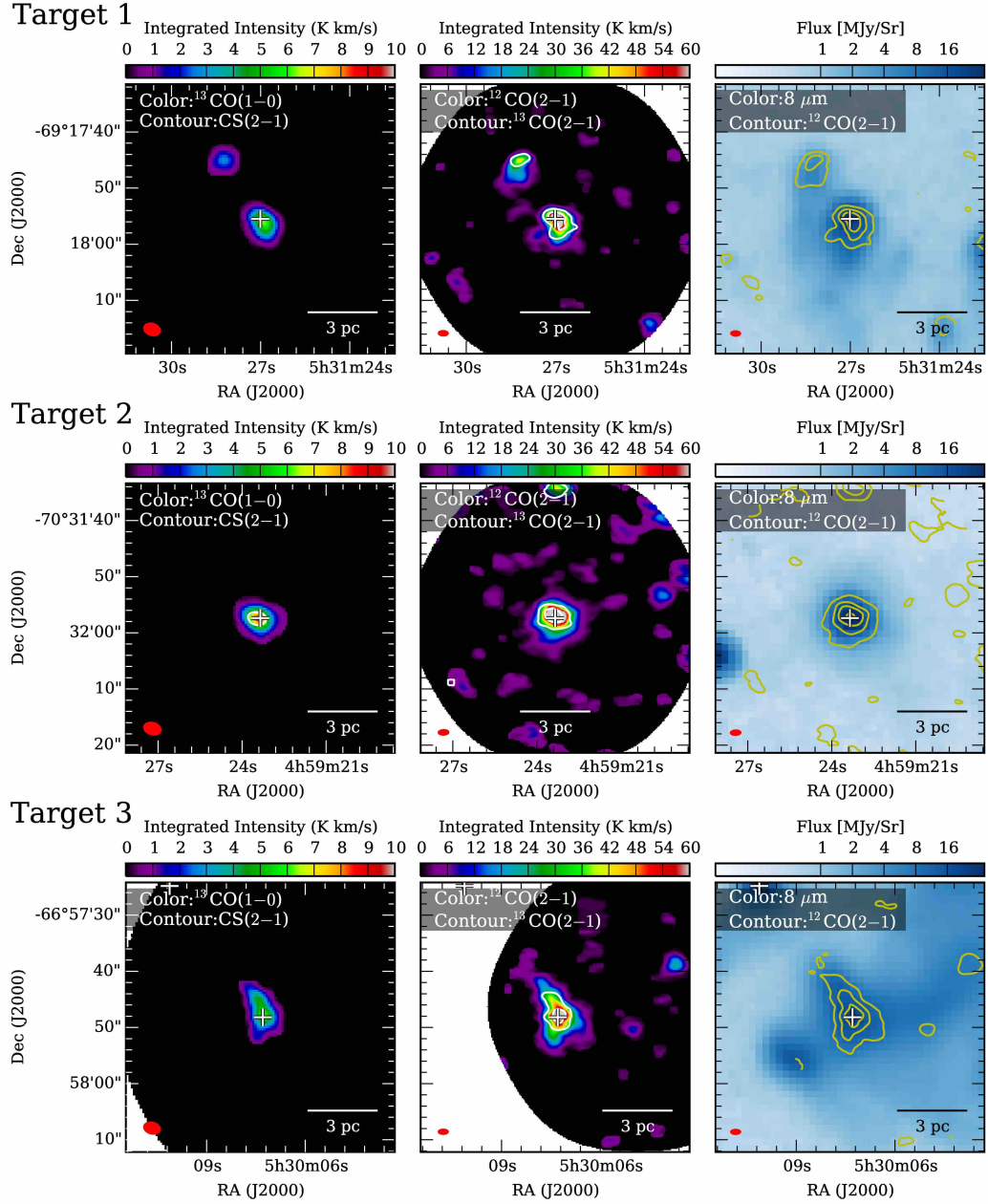}
 \end{center}
\caption{Molecular cloud distributions toward the isolated high-mass YSOs traced by the ALMA observations. White crosses in all panels denote the positions of the YSOs. 
(Left column): ALMA $^{13}$CO ($J$ = 1--0) velocity-integrated intensities are shown in color-scale.  White contours are the CS ($J$ = 2--1) integrated intensities.  The contour levels are 3, 9, 15 and 24$\sigma$ noise level, where 1$\sigma\;$=$\;$0.3$\;$K$\;$km$\,$s$^{-1}$.  The angular resolutions of the $^{13}$CO ($J$ = 1--0) observations are shown at the lower left corners.
(Center column): Color-scale image and white contours show the $^{12}$CO ($J$ = 2--1) and $^{13}$CO ($J$ = 2--1) velocity-integrated intensities, respectively.  The contour levels are 3, 9, 15 and 24$\sigma$ noise level, where 1$\sigma\;$=$\;$1.1$\;$K$\;$km$\,$s$^{-1}$.  The angular resolutions of the $^{12}$CO ($J$ = 2--1) observations are shown at the lower left corners.
(Right column): The $^{12}$CO ($J$ = 2--1) contours are overlaid on the 8 $\mu$m maps obtained with {\it Spitzer} \citep{Meixner06}.  The contour levels are 3, 15, 30 and 60$\sigma$ noise level, where 1$\sigma\;$=$\;$1.6$\;$K$\;$km$\,$s$^{-1}$.  The angular resolutions of the $^{12}$CO ($J$ = 2--1) observations are shown at the lower left corners.
\label{fig:ii_target01-03}}
\end{figure}

\addtocounter{figure}{-1}
\begin{figure}[ht!]
 \begin{center}
  \includegraphics[width=130mm]{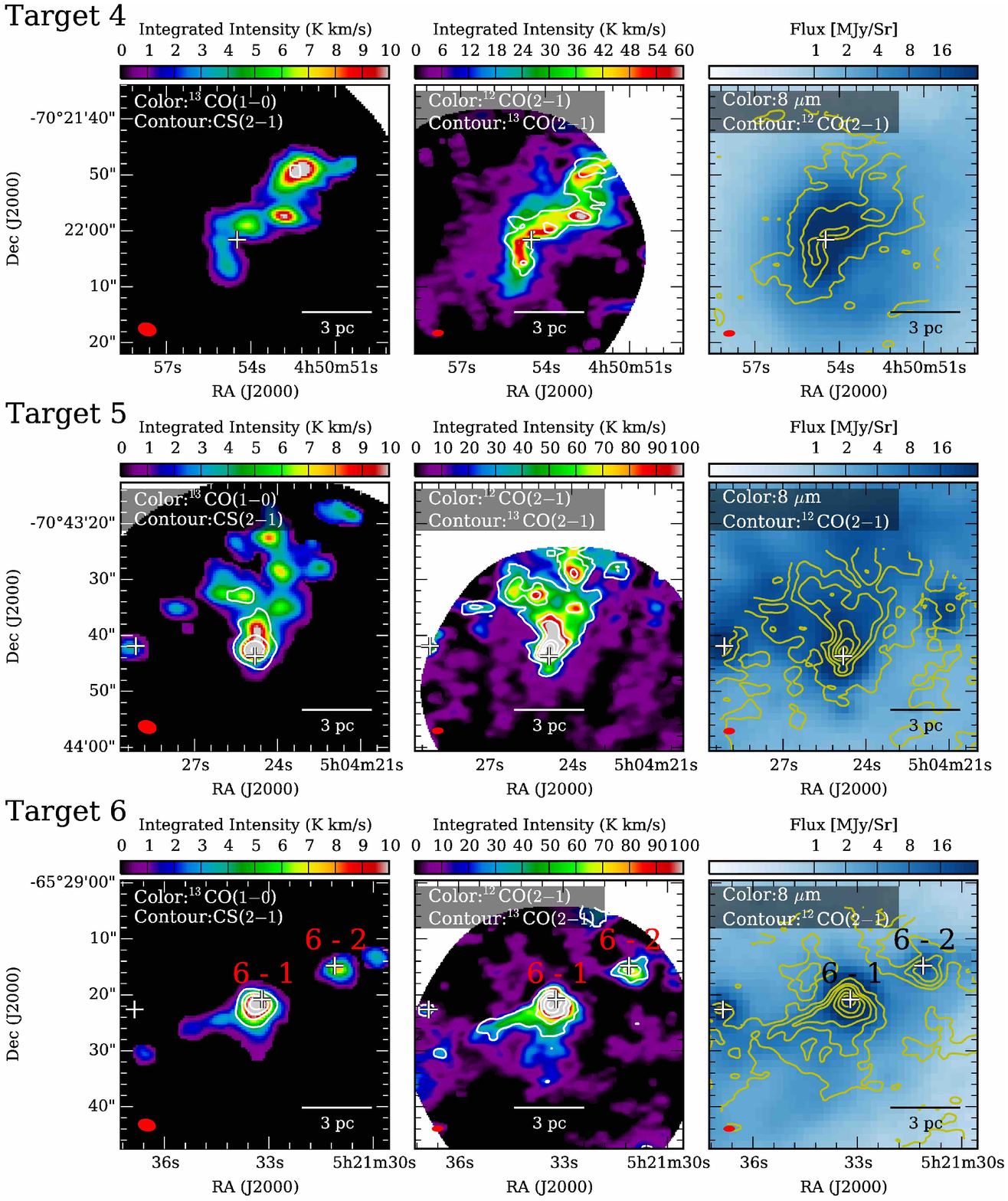}
 \end{center}
\caption{(Continued)
\label{fig:ii_target04-06}}
\end{figure}

\addtocounter{figure}{-1}
\begin{figure}[ht!]
 \begin{center}
  \includegraphics[width=130mm]{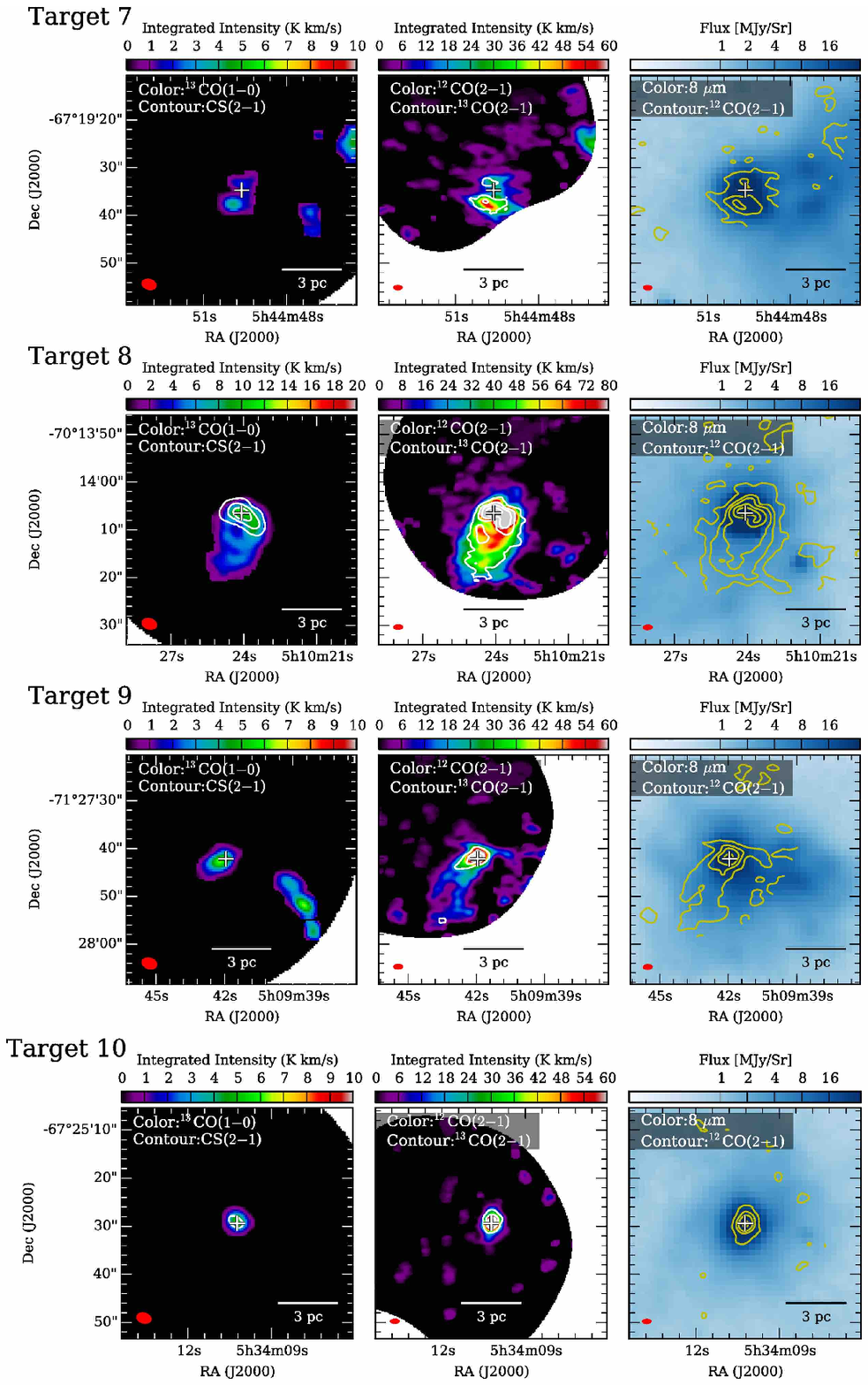}
 \end{center}
\caption{(Continued)
\label{fig:ii_target07-10}}
\end{figure}

\begin{figure}[ht!]
 \begin{center}
  \includegraphics[width=140mm]{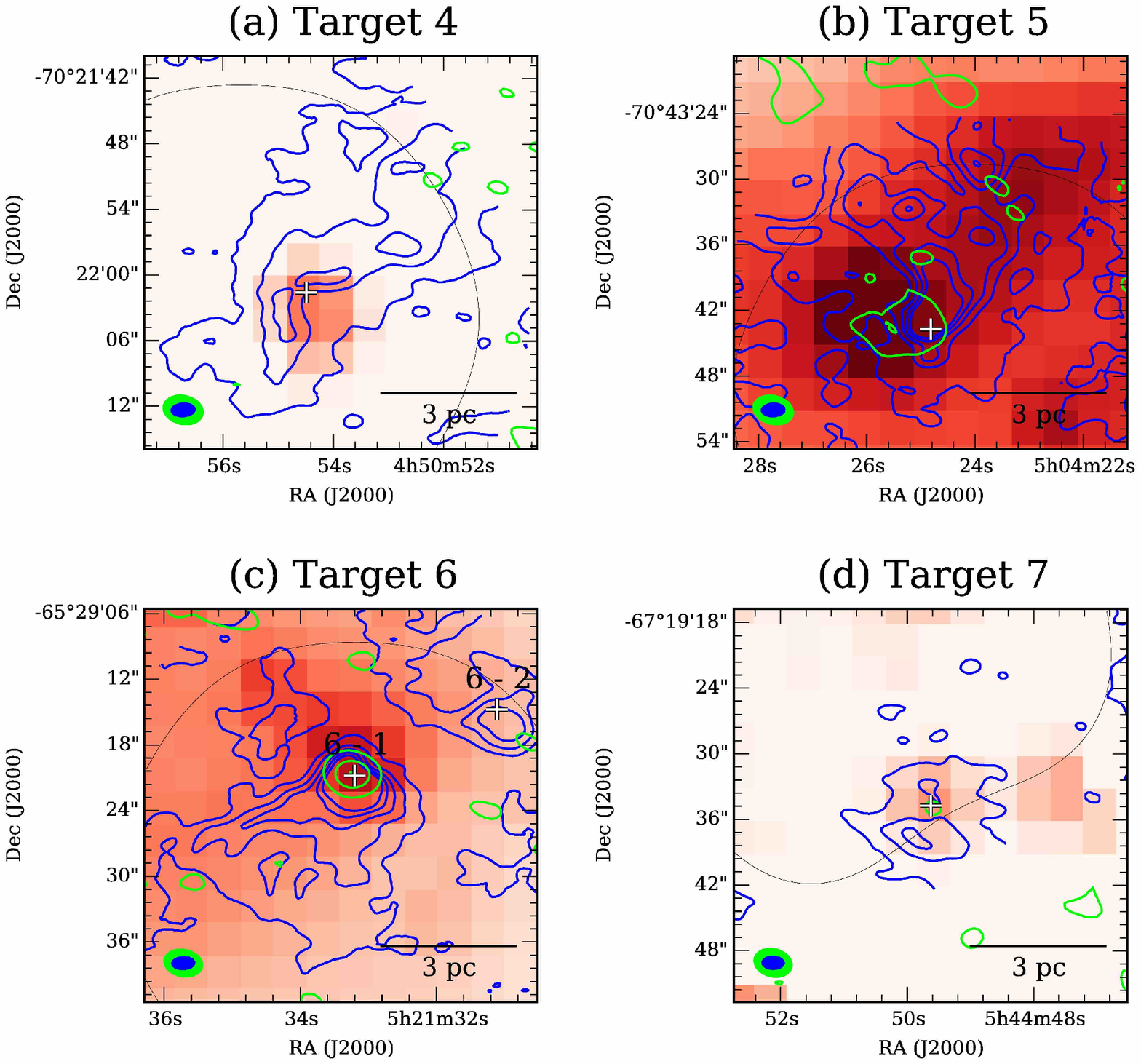}
 \end{center}
\caption{Distributions of molecular gas and ionized gas toward YSOs with a 100$\;$GHz continuum detection (Target 5 and 6) and toward YSOs with the surrounding gas possibly dissipating by the YSOs (Target 4 and 7).  Blue and green contours are the $^{12}$CO ($J$ = 2--1) integrated intensity and the 100$\;$GHz continuum map, respectively.  The contour levels of $^{12}$CO ($J$ = 2--1) are 3, 15, 30 and 60$\sigma$ noise level, where 1$\sigma\;$=$\;$1.6$\;$K$\;$km$\,$s$^{-1}$.  The contour levels of 100$\;$GHz continuum are 3, 9 and 15$\sigma$ noise level, where 1$\sigma\;$=$\;$0.13$\;$mJy$\,$beam$^{-1}$.  Red-color images show the H${\alpha}$ emission \citep{Smith99}.  White crosses show the positions of the YSOs.  Black lines indicate 50\% of the peak sensitivity in the ALMA mosaic.  Blue and green ellipses in the lower left corners in each panel are the angular resolutions of the $^{12}$CO ($J$ = 2--1) and the 100 GHz continuum observations, respectively.  
\label{fig:continuum}}
\end{figure}

\subsection{Velocity structure of associated molecular clouds} \label{subsec:velocity_structure}
\begin{figure}[ht!]
 \begin{center}
  \includegraphics[width=100mm]{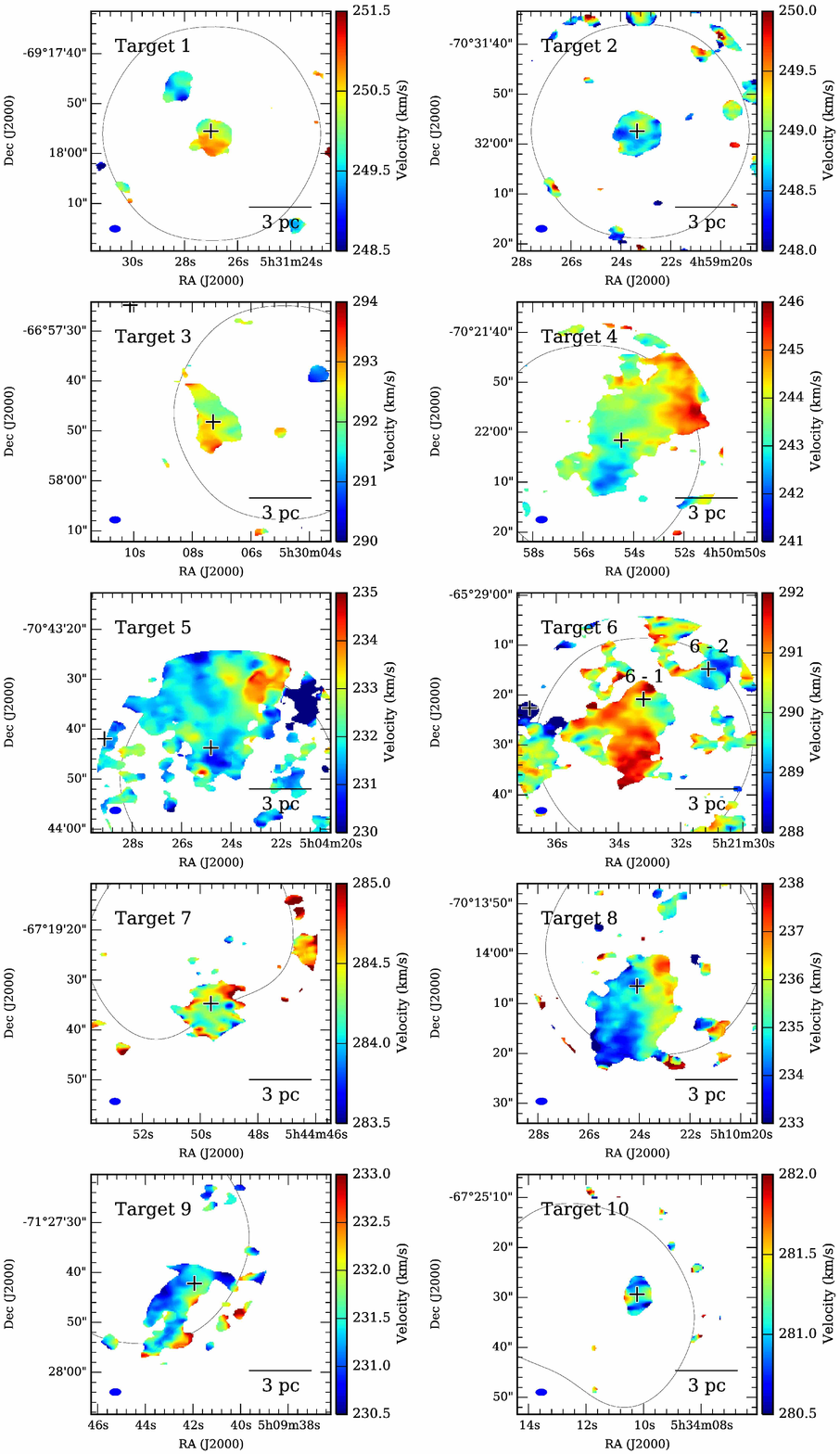}
 \end{center}
\caption{Velocity maps toward the YSOs. 
The first-moment intensity-weighted velocity maps of the $^{12}$CO ($J$ = 2--1) data are shown in color-scale.  Black crosses indicate in each panel the positions of the YSOs.  Black lines indicate 50\% of the peak sensitivity in the ALMA mosaic. The angular resolutions are shown in the lower left corners in each panel.
\label{fig:mom1}}
\end{figure}

\begin{figure}[ht!]
 \begin{center}
  \includegraphics[width=100mm]{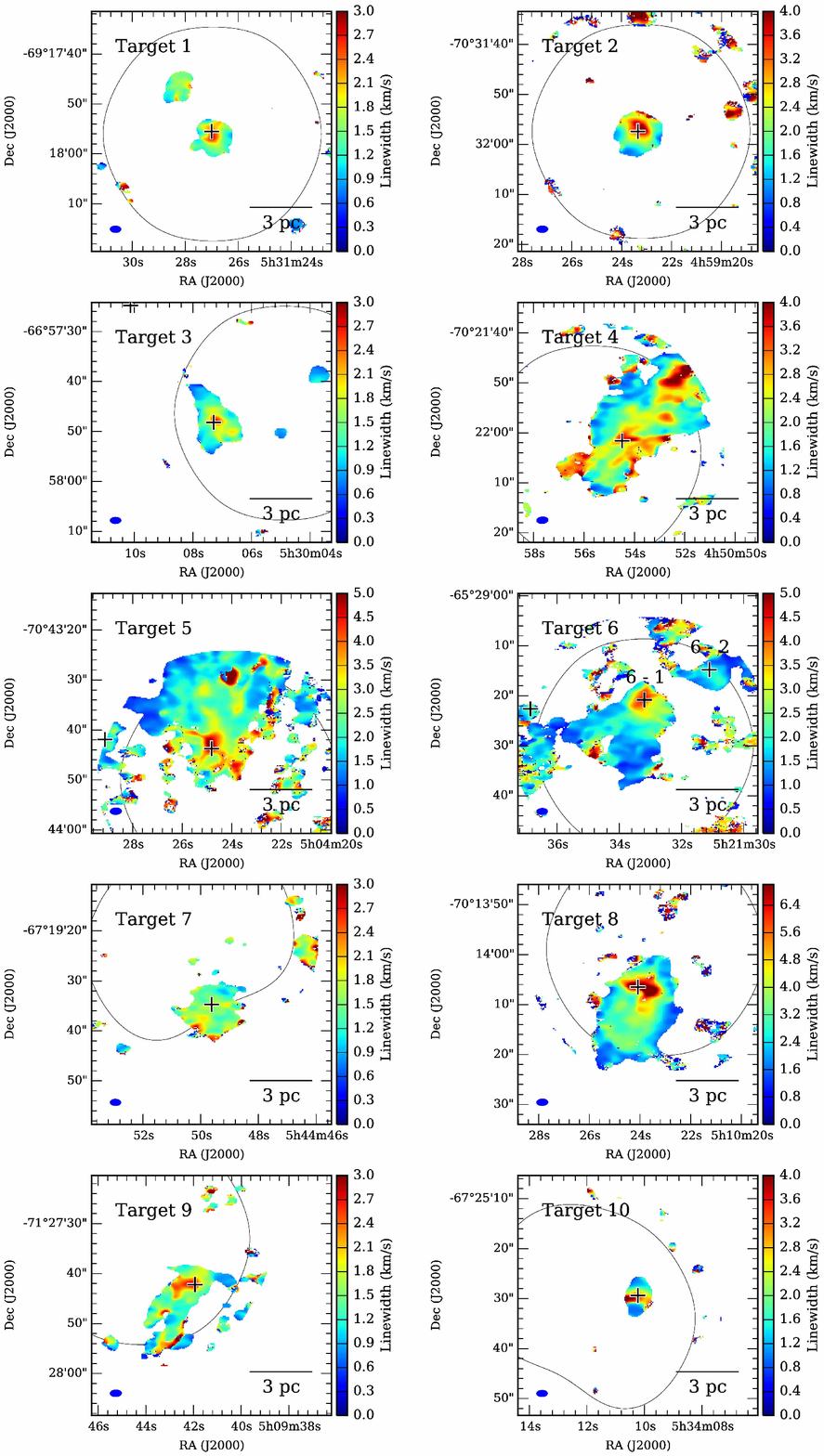}
 \end{center}
\caption{Velocity linewidth maps toward the YSOs. The linewidths (FWHM) of the $^{12}$CO ($J$ = 2--1) data are determined by the single Gaussian fitting to the observed profile in each pixel.  Black crosses indicate in each panel the positions of the YSOs.  Black lines indicate 50\% of the peak sensitivity in the ALMA mosaic. The angular resolutions are shown in the lower left corners in each panel.
\label{fig:line_width}}
\end{figure}

\begin{figure}[ht!]
 \begin{center}
  \includegraphics[width=100mm]{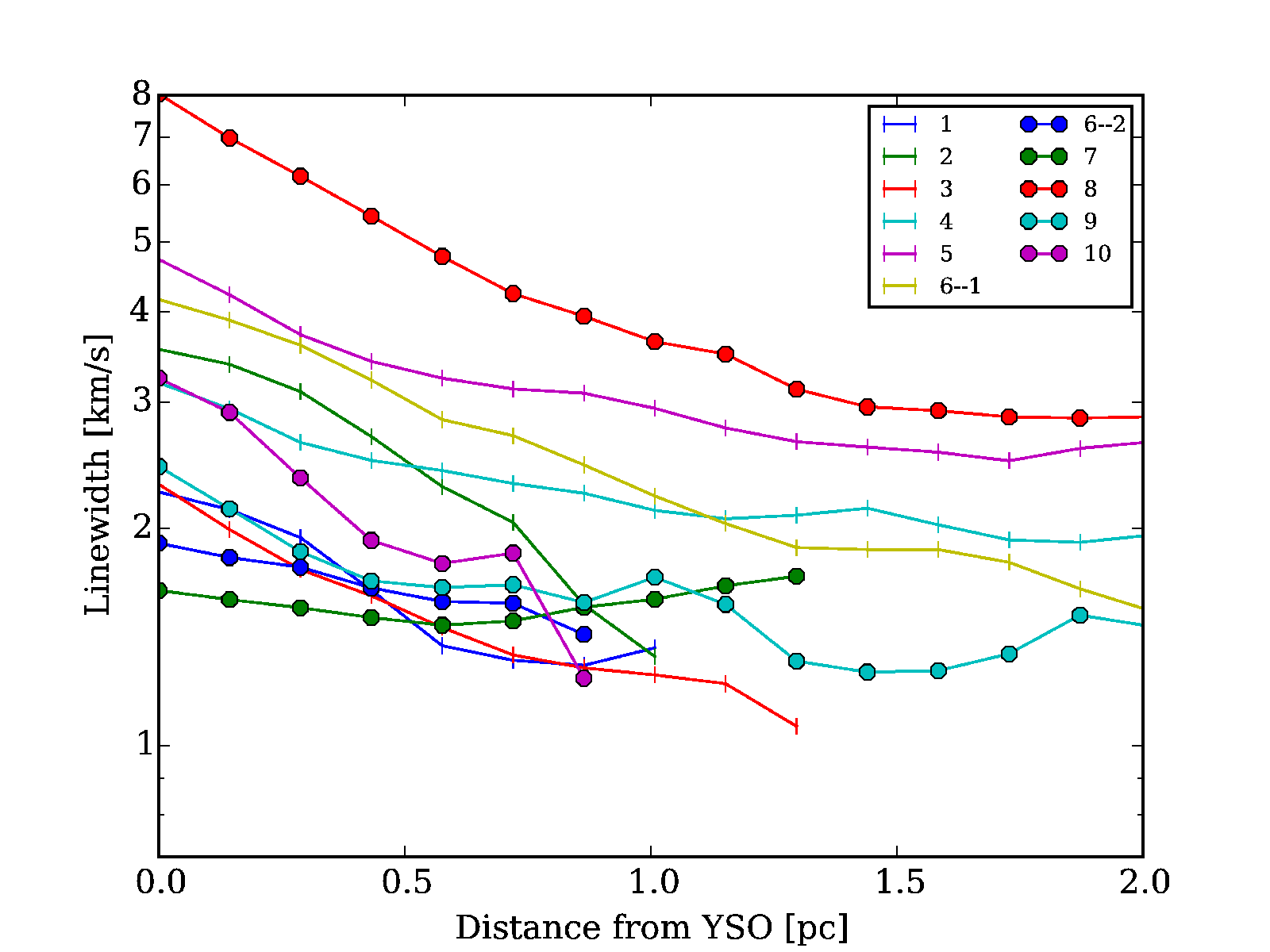}
 \end{center}
\caption{Mean radial distributions of the velocity width centered at each YSO positions made from figure \ref{fig:line_width}.
\label{fig:line_width_profile}}
\end{figure}

\begin{figure}[ht!]
 \begin{center}
  \includegraphics[width=140mm]{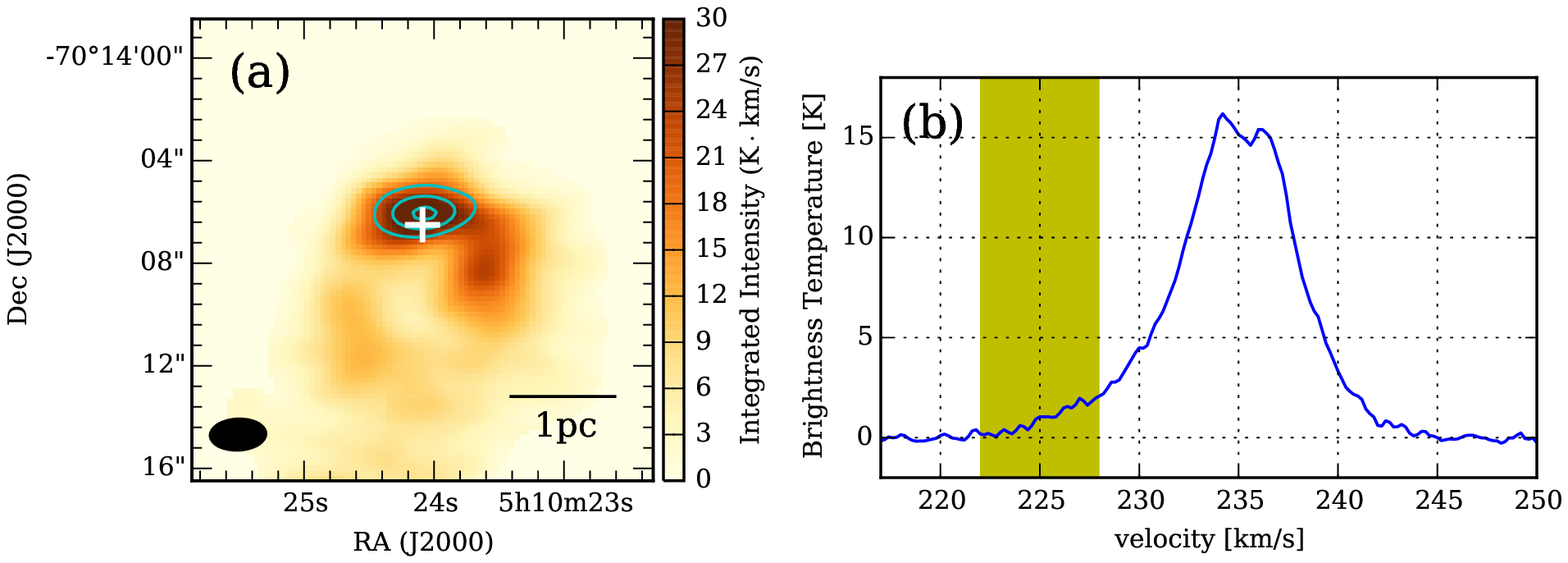}
 \end{center}
\caption{Distribution and velocity profile of the blue-shifted outflow from the YSO Target 8.  (a) Color-scale image shows the velocity-integrated intensity of $^{13}$CO ($J$ = 2--1).  Cyan contours show the $^{12}$CO ($J$ = 2--1) emission integrated over the velocity range shown as the yellow zone in panel (b).  The white cross shows the position of the YSO.  The angular resolution is indicated at the lower left corner as an ellipse.  (b) The blue line shows the averaged $^{12}$CO ($J$ = 2--1) spectrum over the region where the outflow wing is seen.
\label{fig:outflow_map}}
\end{figure}

Figures \ref{fig:mom1} and \ref{fig:line_width} show the kinematic structure of the clouds associated with the YSOs.  The first moment maps show that there is no significant velocity shift toward the compact clouds such as Target 1, 2, and 10.  Some extended clouds show a complex velocity field (Target 5, 6, and 8), and Target 8 shows a velocity shift by 5$\;$km$\,$s$^{-1}$ from North-West to the South-East of the source.  For linewidth maps, the linewidths tend to be larger toward the position of the YSOs.  Figure \ref{fig:line_width_profile} shows the averaged radial distributions of the linewidth centered at the position of the YSOs: we see a gradual increase of the linewidth towards the central YSO, rather than a sharp jump at small radii.  Especially toward Target 8, the linewidth enhancement is obvious toward the source.  Figure \ref{fig:outflow_map} shows the $^{12}$CO ($J$ = 2--1) spectrum toward Target 8, and there is a clear sign of blue-shifted velocity wing due to a molecular outflow.  Even though this is the second most luminous target (see table \ref{tab:sample}), there is no H$\alpha$ emission detected, indicating that this source is at a very early stage of star formation.  This is supported by the fact that free-free emission is undetected toward this source, based on ATCA 3 and 6$\>$cm observations \citep{Stephens13}.  Figure \ref{fig:outflow_map} (a) shows the spatial distribution of the wing emission on the $^{13}$CO ($J$ = 2--1) integrated intensity map.  The wing is very compact with a size of $<$$\;$0.5$\>$pc, located a little bit north of the YSO by 0.1$\>$pc and coincides with the molecular peak.  This fact indicates that the source is very young and the YSO is dissipating the molecular gas south of the YSO, which is consistent with the non-detection of red-shifted outflow wing.  We could not detect outflow wings toward other sources.  The reason may be that Target 8 is the youngest among the sources, having the highest luminosity among the sources without H$\alpha$ or free-free emission, and therefore, the other sources may be more evolved, or the intrinsic outflow activity is lower.  Further high spatial resolution observations would be needed to detect the outflow activities for the other sources.

\subsection{LTE mass } \label{subsec:lte_mass}
The $^{13}$CO ($J$ = 1--0) line intensities give a better estimate of column densities than $^{12}$CO line ones whose optical depth are typically thick; the $^{13}$CO ($J$ = 1--0) line is considered to be optically thin toward typical molecular clouds.  We thus assume the local thermodynamic equilibrium (LTE) to derive the column density from the $^{13}$CO luminosity.  To characterize the gas condensations, we defined a clump as the region where the integrated intensity of $^{13}$CO ($J$ = 2--1) emission is stronger than the 3$\sigma$ level (see figure \ref{fig:ii_target01-03}) around each YSO because $^{13}$CO ($J$ = 2--1) observations have better angular resolutions than $^{13}$CO ($J$ = 1--0) observations.  The physical properties are derived for these clumps.   We derived $^{13}$CO ($J$ = 1--0) column densities, $N$($^{13}$CO) [cm$^{-2}$], by using the following relations by assuming the LTE \citep{Wilson14}: 

\begin{equation}
\label{eq:tau_13co}
\tau_{V}(^{13}\mathrm{CO})=-\mathrm{ln}\left(1-\frac{T_{V}(^{13}\mathrm {CO})/5.3}{1/(\mathrm{exp}[\frac{5.3}{T_{\mathrm{ex}}}]-1)-0.16}\right)
\end{equation}
\begin{equation}
\label{eq:n_13co}
N(^{13}\mathrm{CO})\;=\;3.0\;\times\;10^{14}\;\frac{T_{\mathrm{ex}}\int\;\tau^{13}_{V}dV}{1-\mathrm{exp}[\frac{-5.3}{T_{\mathrm{ex}}}]},
\end{equation}
where $T_{V}(^{13}\mathrm {CO})$ is the $^{13}$CO ($J$ = 1--0) brightness temperature at the velocity.
The excitation temperatures,  $T_{\mathrm{ex}}$, was assumed to be 20$\;$K (e.g., \cite{Nishimura15}).  We used the same abundance ratios of [$^{12}$CO/H$_2$] and [$^{12}$CO/$^{13}$CO] as \citet{Mizuno10} for their N159 study in the LMC as 1.6$\;$$\times$$\;$10$^{-5}$ and 50, respectively, to derive the column density of molecular hydrogen from $N(^{13}$CO) as follows:

\begin{equation}
\label{eq:n_h2}
N(\mathrm{H}_2)=\frac{50}{1.6\;\times\;10^{-5}}\;N(^{13}\mathrm{CO}).
\end{equation}
The masses of the clump, $M_{\mathrm{clump}}$, are calculated from the following equation:
\begin{equation}
\label{eq:n_h2}
M_\mathrm{clump}\;=\;\mu m(\mathrm{H}_2)\;\sum_{}^{} [D^2\Omega N(\mathrm{H}_2)],
\end{equation}
where $\mu$ is the mean molecular weight per hydrogen molecule, $m$(H$_2$) is the H$_2$ molecular mass, $D$ is the distance to the object of 50$\>$kpc, and $\Omega$ is the solid angle of a pixel element.  The masses of the clumps associated with the isolated high-mass YSOs range from $\sim$$\;$300 to $\sim$$\;$6,000$\;$$M_{\odot}$ (see table \ref{tab:cloud_property}).  

\begin{table}[ht!]
  \tbl{clump properties}{%
  \begin{tabular}{lccccccccccc}
      \hline
Target & \multicolumn{4}{c}{$T_{\mathrm{peak}}$} & $R_{\mathrm{deconv}}$ & ${\Delta}V_{\mathrm{clump}}$ & $M_{\mathrm{vir}}$ & $M_{\mathrm{LTE}}$ & $M_{\mathrm{LTE}}$ / $M_{\mathrm{vir}}$ & $N$(H$_2$)\\
      \cline{2-5}
No. & $^{12}$CO & $^{13}$CO & $^{13}$CO & CS &  &  &  &  &  & \\
 & ($J$ = 2--1) & ($J$ = 2--1) & ($J$ = 1--0) & ($J$ = 2--1) &  &  &  &  &  & \\
 & [K] & [K] & [K] & [K] & [pc] & [km$\,$s$^{-1}$] & [$\times$$\;$10$^{2}$$\;$$M_{\odot}$] & [ $\times$$\;$10$^{2}$$\;$$M_{\odot}$] &  & [$\times$$\;$10$^{22}$$\;$cm$^{-2}$]\\
 & (1) & (2) & (3) & (4) & (5) & (6) & (7) & (8) & (9) & (10)\\
      \hline
1 & 29.7  & 10.9  & 5.0  & ... & 0.6  & 1.4  & 2.2  & 6.0 & 2.7  & 1.8 \\
2 & 26.0  & 7.7  & 4.0  & 0.94 & 0.6  & 2.1  & 5.2  & 7.2 & 1.4  & 2.2\\
3 & 30.4  & 9.5  & 5.1  & ... & 0.6  & 1.6  & 3.0  & 6.4 & 2.1  & 1.9\\
4 & 24.0  & 9.7  & 7.2  & ... & 1.6  & 3.1  & 28.9  & 48 & 1.7  & 2.5\\
5 & 40.5  & 13.9  & 4.8  & 0.84 & 2.0  & 4.1  & 62.9  & 65 & 1.0  & 2.1 \\
6--1 & 55.2  & 14.3  & 6.4  & 1.3 & 1.2  & 4.0  & 36.6  & 28 & 0.8  & 2.4\\
6--2 & 30.9  & 14.0  & 5.5  & ... & 0.6  & 1.8  & 3.5  & 5.6 & 1.6  & 1.9\\
7 & 28.3  & 8.4  & 3.4  & ... & 0.6  & 2.1  & 5.0  & 2.8 & 0.6  & 1.0\\
8 & 88.0  & 9.6  & 4.1  & 0.74 & 1.6  & 4.6  & 63.5  & 45 & 0.7  & 2.5\\
9 & 43.3  & 10.2  & 3.9  & ... & 0.5  & 1.9  & 3.6  & 4.6 & 1.3  & 1.9\\
10 & 31.0  & 11.7  & 4.0  & ... & 0.5  & 2.5  & 5.8  & 3.6 & 0.6  & 1.5\\
    \hline
    \end{tabular}}\label{tab:cloud_property}
\begin{tabnote}
Col.(1)--(4): Brightness temperature derived by fitting the profile with a single Gaussian function at the peak intensity position of the source.  The spectra are measured at the angular resolution of $\timeform{1".9}$$\;$$\times$$\;$$\timeform{1".1}$ for $^{12}$CO, $^{13}$CO ($J$ = 2--1) and $\timeform{3".2}$$\;$$\times$$\;$$\timeform{2".4}$ for $^{13}$CO ($J$ = 1--0) and CS ($J$ = 2--1). \\
(5): Effective radius of a circle having the same area as that of the region above the 3 $\sigma$ level on the $^{13}$CO ($J$ = 2--1) integrated intensity map. The radius is deconvolved by the Band 6 beam size, using the geometric mean between the major and minor axis. \\
Col.(6): Linewidth (FWHM) derived by a single Gaussian fitting to $^{13}$CO ($J$ = 1--0) spectrum at the intensity peak. \\
Col.(7): Virial mass derived from $R_{\rm deconv}$ (Col. (5)) and $\Delta V_{\rm clump}$ (Col. (6)). \\
Col.(8): LTE mass derived from the $^{13}$CO ($J$ = 1--0) data (see the text).  \\
Col.(10): Averaged H$_{2}$ column density in the area above the 3 $\sigma$ level of $^{13}$CO ($J$ = 2--1) integrated intensity map. \\
\end{tabnote}
\end{table}

\subsection{Virial mass } \label{subsec:virial_analysis}
The mass can be also estimated from the dynamical information of the size of the clumps and the linewidth of CO lines by assuming the virial equilibrium.  We derived the virial masses from the $^{13}$CO ($J$ = 1--0) line using the procedure described by \citet{Fujii14} as follows. The virial mass is estimated as
\begin{equation}
\label{eq:virial}
M_{\mathrm{vir}}\; [$\MO$]=190\;{\Delta}V^{2}_{\mathrm{clump}}\;[\mathrm{km}\;\mathrm{s}^{-1}]\;R_{\mathrm{deconv}}\;[\mathrm{pc}],
\end{equation}
assuming the clumps are spherical with density profiles of $\rho$  $\propto$ $r^{-1}$, where $\rho$ is the number density, and $r$ is the distance from the cloud center \citep{MacLaren88}.  Deconvolved clump sizes, $R_{\mathrm{deconv}}$, are defined as [$R^2_{\mathrm{observed}}$ - ($\theta_{\mathrm{HPBW}}$/2)$^2$]$^{1/2}$.  We calculated the effective radius as ($A/\pi)^{1/2}$ where $A$ is the observed total cloud surface area.  The virial ratio is derived by dividing the virial mass by the LTE mass.  The virial ratio of the clumps associated with the isolated high-mass YSOs range from 0.6 to 2.7 (see table \ref{tab:cloud_property}).

\subsection{Excitation analyses using multiple transitions} \label{subsec:lvg}
\begin{figure}[ht!]
 \begin{center}
  \includegraphics[width=100mm]{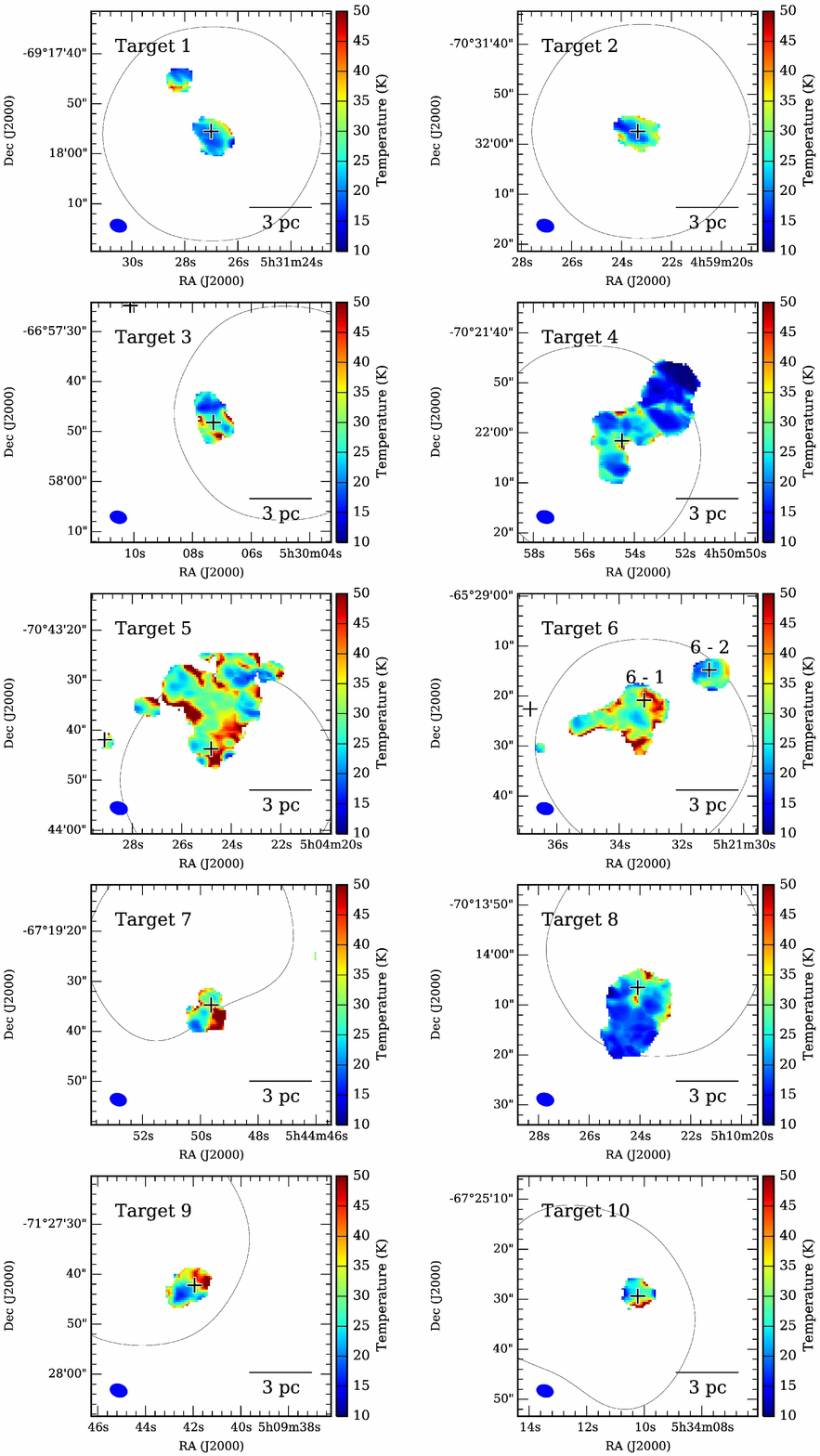}
 \end{center}
\caption{Maps of the gas kinetic temperature toward the YSOs calculated by the LVG analyses (see the text).  Black crosses indicate in each panel the positions of the YSOs.  Black lines indicate 50\% of the peak sensitivity in the ALMA mosaic.  The angular resolutions are shown in the lower left corners in each panel.
\label{fig:temperature}}
\end{figure}

\begin{figure}[ht!]
 \begin{center}
  \includegraphics[width=100mm]{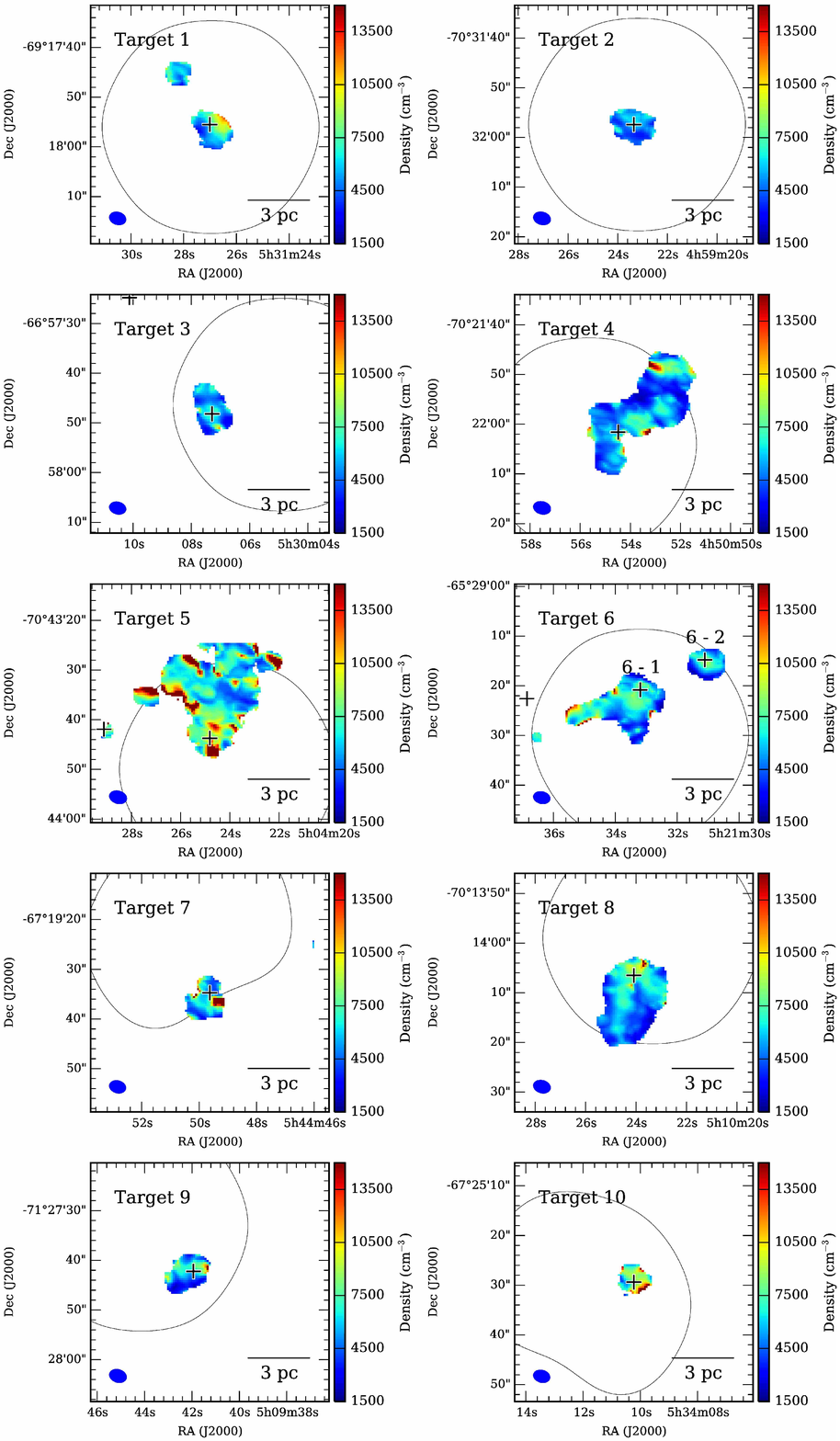}
 \end{center}
\caption{H$_2$ density maps toward the YSOs calculated by the LVG analyses (see the text).  Black crosses indicate in each panel the positions of the YSOs.  Black lines indicate 50\% of the peak sensitivity in the ALMA mosaic.  The angular resolutions are shown in the lower left corners in each panel.
\label{fig:density}}
\end{figure}

The kinetic temperature and number density of the molecular gas can be estimated by observing multiple lines having different critical densities and optical depths.  In order to derive the properties, we performed a large velocity gradient analysis (LVG analysis; \cite{Goldreich74}; \cite{Scoville74}) for our CO line observations.  The assumption of the LVG analysis is that the molecular cloud is spherically symmetric with uniform density and temperature, and having a spherically symmetric velocity gradient proportional to the radius.  The model requires three independent parameters: the kinetic temperature, $T_{\mathrm{kin}}$, the density of molecular hydrogen, $n$(H$_2$), and the fractional abundance of CO divided by the velocity gradient in the cloud, X(CO)/($dv/dr$).  We use the abundance ratios of [$^{12}$CO/H$_2$] and  [$^{12}$CO/$^{13}$CO] described in section \ref{subsec:lte_mass}.  The mean velocity gradient is estimated as $dv/dr$ (km$\,$s$^{-1}$$\,$/$\,$pc) = ${\Delta}V_{\mathrm{clump}}\;/\;2R_{\mathrm{deconv}}$, and we adopt $dv/dr$ = 2 from the typical size and linewidth of the clumps. According to the analysis using the same molecular lines by \citet{Nishimura15}, the derived density is inversely proportional to the square root of the X(CO)/($dv/dr$).\\
\ \ The distributions of the kinetic temperatures and densities are shown in figures \ref{fig:temperature} and \ref{fig:density}, and the averaged values are described in table \ref{tab:lvg}.  The density is in the range of 5.4--8.8$\;$$\times$$\;$10$^3$$\;$cm$^{-3}$, and the kinetic temperatures are 19--30 K, which is consistent with the dust temperatures derived by \citet{Seale14}.  We note that the kinetic temperatures do not show clear local enhancement toward the position of the YSOs.  This implies that the heating by the newly formed YSOs is not yet significant.

\begin{table}[ht!]
  \tbl{Results of the LVG analyses toward each YSO.}{%
  \begin{tabular}{lccc}
      \hline
Target No. & $n$(H$_2$) & $T_{\mathrm{kin}}$ & $T_{\mathrm{FIR}}$ \\
& [$\times$$\;$10$^3$ cm$^{-3}$] & [K] & [K] \\
 & (1) & (2) & (3) \\
      \hline
1 & 6.8  & 21  & 16.8\\
2 & 5.4  & 22  & 25.5\\
3 & 5.7  & 26  & 16.3\\
4 & ...  & ... & 21.8\\
5 & 7.1  & 30  & 25.7\\
6--1 & 6.3  & 30  & 28.9\\
6--2 & 6.3  & 25  & ...\\
7 & ...  & ... & ...\\
8 & 5.8  & 19  & 23.6\\
9 & 6.5  & 29  & 22\\
10 & 8.8  & 28  & 26\\
      \hline
    \end{tabular}}\label{tab:lvg}
\begin{tabnote}
Col.(1): Averaged H$_2$ volume density derived from figure \ref{fig:density}.  Col.(2): Averaged gas kinematic temperature derived form figure \ref{fig:temperature}.  Col.(3): Dust temperature derived by graybody fitting to the far-IR data \citep{Seale14}.  The $^{13}$CO ($J$ = 2--1) emission is not detected toward the positions of Target 4 and 7, and thus we cannot derive the physical properties.  There are no FIR temperatures available for Target 6--2 and 7 in \citet{Seale14}.
\end{tabnote}
\end{table}

\section{Discussion} \label{sec:discussion}

\subsection{Clump dynamics} \label{subsec:clump_dynamics}
Table \ref{tab:cloud_property} shows that the virial ratio, $M_{\mathrm{LTE}}$/$M_{\mathrm{vir}}$, is distributed around unity.  We assumed the abundance ratio of $^{13}$CO to be a typical value in the LMC as described in section \ref{subsec:lvg}, which is $\sim$$\;$1/3 of the Galactic value.  If the abundance ratio is correct, the virial ratios indicate that the clumps are roughly gravitationally bound.  Each cloud has a high-mass YSO inside so that the observation of gravitational boundedness seems to be appropriate, although gas dissipation may be starting for some of the clumps.

\subsection{Photodissociation of CO} \label{subsec:photodissociation}
The densities of the CO clumps range from 5.4 to 8.8$\;$$\times$$\;$10$^3$$\;$cm$^{-3}$ based on the excitation analysis (see table \ref{tab:lvg}).  We also estimated the densities from the virial mass for the circularly shaped clouds, i.e., Target 1, 2, 3, and 10, under the assumption of the spherical morphology; the densities are estimated to be 3,500, 7,600, 3,800, and 16,500$\;$cm$^{-3}$, respectively.  These values are roughly consistent with the densities derived from the excitation analysis, and thus the density estimation seems to be roughly appropriate.  These densities are higher than the typical ones toward GMCs in the Galaxy by a factor of a few or more if we observe them in the same probe; for Orion molecular cloud, the average density is estimated to be 2$\;$$\times$$\;$10$^3$$\;$cm$^{-3}$ (e.g., \cite{Nishimura15}).  This may be due to the fact that in the LMC, the gas-to-dust ratio is higher by a factor of $\sim\;$2--3 and that the UV radiation is stronger than in the Galaxy, and thus CO can be more strongly photodissociated than in the Galaxy, which results in the photodissociation of lower density gas in the LMC.  The ALMA observations of N83C in the SMC showed that the densities measured in $^{12}$CO are $\sim$$\;$10$^4$$\;$cm$^{-3}$, and the kinetic temperature is estimated to be 30--50$\;$K throughout the cloud \citep{Muraoka17}, and they argued that this is due to the effect of photodissociation in a low metallicity environment.  Although the metallicity at the SMC is below the LMC (0.2$\;Z_{\odot}$; \cite{Dufour84}; \cite{Kurt99}; \cite{Pagel03}), this fact indicates that CO observations can miss molecular mass in the low-metallicity/strong-UV environments in the LMC unlike the case in the Galaxy.  In this case, [C$\;${\sc i}] (1--0) may be another proper probe for the molecular mass (e.g., \cite{Papadopoulos04}).

\subsection{Compact clumps associated with the isolated high-mass YSOs} \label{subsec:compact_clump}
The masses of the clumps associated with the isolated high-mass YSOs range from $\sim$$\;$300 to $\sim$$\;$6,000$\;$$M_{\odot}$ and the radii from 0.5 to 2.0$\;$pc (see table \ref{tab:cloud_property}), which is similar to those of typical nearby dark clouds where low-mass stars are forming, and much smaller than the typical high-mass star-forming clouds in the Galaxy, such as the Orion molecular cloud, which has a mass of $\sim$$\;$10$^5$$\;$$M_{\odot}$ \citep{Nishimura15}.  The NANTEN and Mopra observations show that these clumps are not associated with extensive molecular gas.  These facts indicate that the isolated high-mass YSOs are formed in compact and low-mass clumps.  Most of the present target YSOs are located at the peak of the clumps, indicating that the YSOs were formed in the associated clumps, and the significant dissipation of the gas has not started yet toward the clumps.  From the LVG analysis in section \ref{subsec:lvg}, we found that there is no kinetic temperature enhancement toward the YSOs, which is consistent with the fact that the YSO is very young and the effect of the YSO to the natal gas is not significant.

\subsection{Star Formation Efficiency} \label{subsec:sfe}
\begin{figure}[ht!]
 \begin{center}
  \includegraphics[width=100mm]{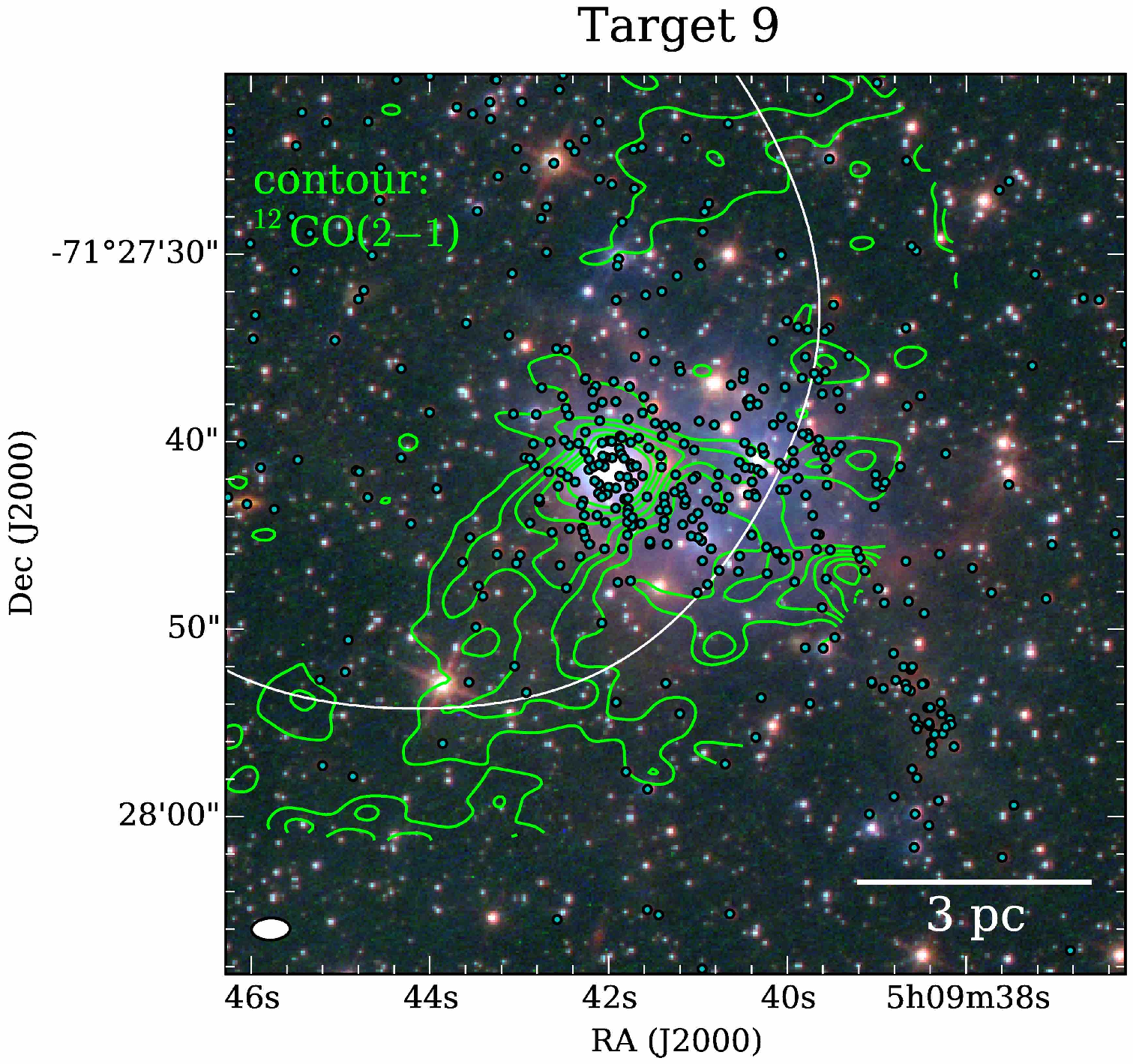}
 \end{center}
\caption{Green contours show the $^{12}$CO ($J$ = 2--1) integrated intensity toward Target 9.  The contour levels are 2.7, 9, 15 and 24$\sigma$ noise level, where 1$\sigma\;$=$\;$1.1$\;$K$\;$km$\,$s$^{-1}$.  The angular resolution is indicated at the lower-left corner as an ellipse.  RGB: F160W (H-band), F814W (I-band), and F555W (V-band).  White line indicates 50\% of the peak sensitivity in the ALMA mosaic.  Cyan filled circles are the positions of the identified YSOs by \citet{Stephens17}.
\label{fig:hst_map}}
\end{figure}

As described in the previous section, the isolated high-mass YSOs in this study appear to have been formed in compact and low-mass molecular clumps, which are much smaller than the typical GMCs.  In this section, we estimate the star formation efficiency of these clumps.  If we assume a normal IMF, the formation of a single high-mass star should be associated with the formation of a number of low-mass stars.  With {\it Spitzer} observations alone, it is hard to identify the low-mass YSOs, but we have to take the mass of the low-mass YSOs into account to estimate the star formation efficiency.  One of the targets, Target 9, was observed by {\it HST} \citep{Stephens17}, and in figure \ref{fig:hst_map} we show the comparison with our CO distribution.  There are a number of low-mass YSOs around the high-mass YSO, and there is a high-concentration of the low-mass YSOs toward the CO clump.  \citet{Weidner13} studied the correlation between the star cluster mass, $M_{\mathrm{ecl}}$, and the mass of the most high-mass star in a star cluster, $m_{\mathrm{max}}$.  {\it HST} observations by \citet{Stephens17} found that relatively isolated high-mass YSOs generally follow the $m_{\mathrm{max}}$--$M_{\mathrm{ecl}}$ relation.  If we use this correlation, the masses of the star cluster associated with the isolated high-mass YSOs range from 110 to 310$\;$$M_{\odot}$ (see table \ref{tab:sfe}).  From this information, we calculated the star formation efficiency with the following equation:

\begin{equation}
\label{eq:A5}
SFE = \frac{M_{\mathrm{ecl}}}{M_{\mathrm{ecl}} + M_{\mathrm{cloud}}},
\end{equation}
where $M_{\mathrm{ecl}}$ is the mass of the star cluster derived from the $m_{\mathrm{max}}$--$M_{\mathrm{ecl}}$ relation \citep{Weidner13} and $M_{\mathrm{cloud}}$ is the LTE mass derived from $^{13}$CO ($J$ = 1--0).  The star formation efficiencies are shown in table \ref{tab:sfe} and they range from 4 to 43\%.  We note that this SFE is an upper limit because a part of the natal clouds may have been already dissipated away and because some molecular gas may be CO-dark (see discussion in section \ref{subsec:photodissociation}).  The efficiencies are high especially for the compact clumps, such as for Target 1, 2, 3, and 10.  Although the high efficiency is seen toward Target 8, the gas dissipation seems to be significantly on-going for this clump (see section \ref{subsec:velocity_structure}), which makes the mass of the associated gas small.  In the case of the Orion molecular cloud, the efficiency was estimated to be $\sim$$\;$4.5\% \citep{Nishimura15}, which means that the star formation is efficiently on-going in the compact molecular clumps we observed in the present paper.

\begin{table}
  \tbl{$SFE$ of the molecular clouds associated with the YSOs.}{%
  \begin{tabular}{cccc}
      \hline
Target No. & \ $m_{\mathrm{max}}$\footnotemark[$*$] & \ $M_{\mathrm{ecl}}$\footnotemark[$\dag$] & \ $SFE$  \\
& [$M_{\odot}$] & [$M_{\odot}$] & [\%] \\
      \hline
1 & 10 & 110 & 16 \\
2 & 13 & 180 & 20 \\
3 & 13 & 180 & 22 \\
4 & 14 & 210 & 4 \\
5 & 17 & 310 & 5 \\
6--1 & 17 & 310 & 10 \\
6--2 & 13 & 180 & 24 \\
7 & 14 & 210 & 43 \\
8 & 16 & 280 & 6 \\
9 & 15 & 240 & 34 \\
10 & 15 & 240 & 40 \\
      \hline
    \end{tabular}}\label{tab:sfe}
\begin{tabnote}
\footnotemark[$*$] Mass of the YSOs derived in section \ref{sec:observations}.\\
\footnotemark[$\dag$] Embedded cluster mass ($M_{\mathrm{ecl}}$) with an assumption of  the typical m$_{\mathrm{max}}$--$M_{\mathrm{ecl}}$ relation \citep{Weidner13}.
\end{tabnote}
\end{table}

\subsection{Enhancement of velocity width toward YSOs} \label{subsec:velocity_width}
As described in section \ref{subsec:velocity_structure}, linewidths increase toward the positions of the YSOs or the peak of the CO intensities.  Here, we discuss the cause of the linewidth enhancement toward the YSOs, and in particular whether it is due to the YSO activity or to the clump properties prior to star formation.  The outflow from YSOs can inject momentum into the natal molecular cloud, which results in the enhancement of the linewidth of molecular clouds (e.g., \cite{Saito07}).  In the case of Target 8, the outflow wing is the apparent cause of the linewidth enhancement.  For the other sources, apparent wing feature cannot be observed at our spatial resolution and sensitivity, which could mean that the current outflow activities for these YSOs are not strong enough to be detected.\\
\ \ Here, we assume that the linewidth enhancement is $\sim\,$1.5$\,$km$\,$s$^{-1}$ at the peak and located within $\sim\;$1$\;$pc of the YSO peak, which is typical for the larger clumps in figure \ref{fig:line_width_profile}.  The central $\sim\;$1$\;$pc corresponds to a mass of $\sim\,$1,500$\,$$M_{\odot}$ by assuming the average column density of 2$\,\times\,$10$^{22}$$\;$cm$^{-2}$.  If we assume that the linewidth enhancement is caused by the momentum input by the outflow, the momentum is calculated to be $\sim\,$1,100$\,M_{\odot}$$\,$km$\,$s$^{-1}$ (=$\,$1/2$\,\times\,$1,500$\,M_{\odot}\,\times\,$1.5$\,$km$\,$s$^{-1}$) under the assumption that the linewidth enhancement is linearly increasing toward the peak and that the column density is inversely proportional to the distance from the center.  Actually, the total momentum contributing to the linewidth enhancement, the difference from 2.2$\,$km$\,$s$^{-1}$ at 1$\,$pc, of Target 6--1 is measured to be $\sim\,$2,200$\,M_{\odot}$$\,$km$\,$s$^{-1}$ by summing up all the excess momenta within 1 pc of the YSO where the average column density is 4$\,\times\,$10$^{22}$$\;$cm$^{-2}$.  Nevertheless, the momentum that can be supplied by outflows is estimated to be 1--300$\,M_{\odot}\,$km$\,$s$^{-1}$ for young stars with a luminosity of 10$^{2}$--10$^{4}$$\,$$L_{\odot}$ (e.g., \cite{Zhang05}).  This means that, even if the outflow has a momentum of the maximum of 300$\;$$M_{\odot}$$\,$km$\,$s$^{-1}$ and all the outflow momentum turns into that of the clump, the momentum of the outflow may not contribute to the linewidth enhancement fully.  Therefore, there is a high possibility that the high-mass YSOs were formed in the turbulent clump region prior to the star formation.  We discuss the interaction between multiple velocity gas components as a mechanism to realize the turbulent environment in the next section.\\

\subsection{The cause of the high-mass star formation in an isolated environment} \label{subsec:ccc}

\begin{figure}[ht!]
 \begin{center}
  \includegraphics[width=100mm]{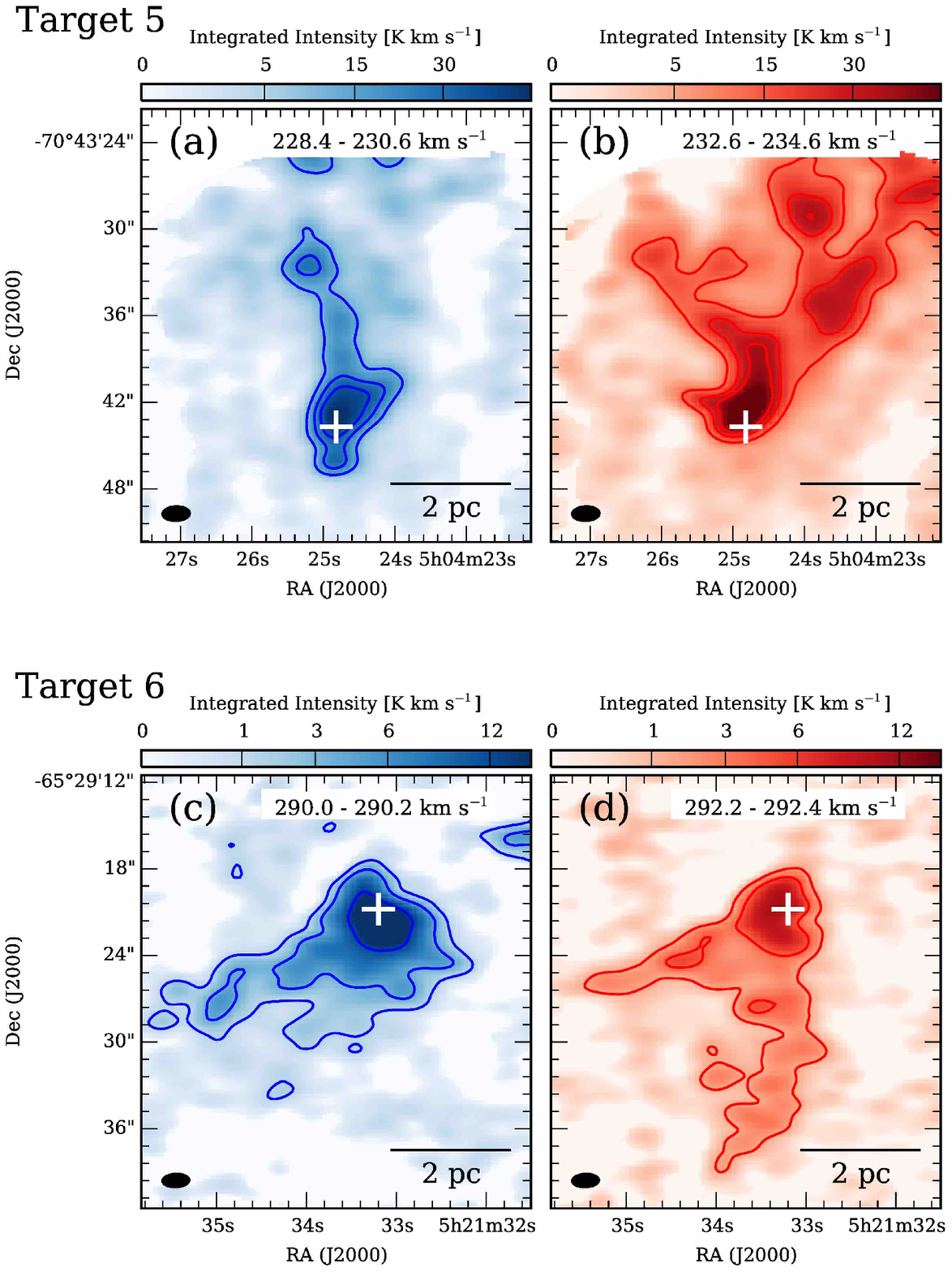}
 \end{center}
\caption{$^{12}$CO ($J$ = 2--1) intensity maps toward Target 5 integrated over velocity ranges of 228.4--230.6 (a) and 232.6-- 234.6 (b) km$\,$s$^{-1}$, respectively.  $^{12}$CO ($J$ = 2--1) intensity maps toward Target 6 integrated over velocity ranges of 290.0--290.2 (c) and 292.2--292.4 (d) km$\,$s$^{-1}$, respectively.  White crosses indicate in each panel the positions of the YSOs.  The angular resolutions are shown in the lower left corners in each panel.
\label{fig:channel_map}}
\end{figure}

\begin{figure}[ht!]
 \begin{center}
  \includegraphics[width=120mm]{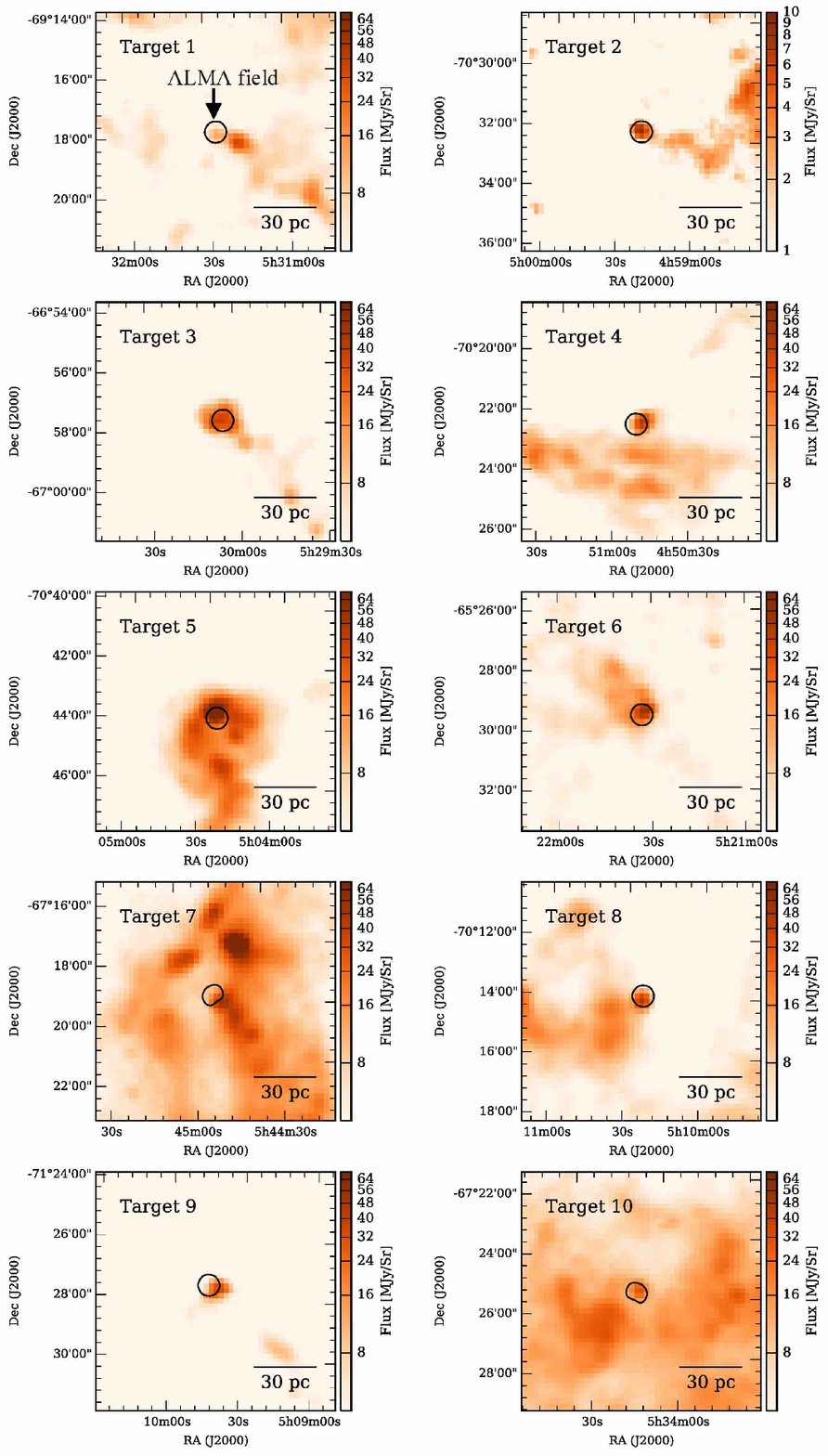}
 \end{center}
\caption{Large-scale views of the dust continuum distribution toward the YSOs. Color-scale shows the {\it Herschel} 350${\mu}m$ images. Black lines indicate 50\% of the peak sensitivity in the ALMA mosaic.
\label{fig:350um}}
\end{figure}

The importance of cloud collisions as a mechanism of the formation of high-mass stars has recently been discussed by many observational and theoretical studies.  Recent mm-submm observations revealed supersonic collisions between multiple molecular clouds having different velocities toward the super star clusters in the Galaxy, e.g., Westerlund$\,$2, RCW$\,$38, and NGC$\,$3603, and the observational studies suggested that the collision is followed by strong shock compression of the molecular gas that leads to the formation of massive clusters in less than 1$\>$Myr (see \cite{Furukawa09}; \cite{Ohama10}; Fukui et al. \yearcite{Fukui14}, \yearcite{Fukui16}).  Recent ALMA observations in the LMC also revealed highly filamentary molecular structure, which is explained by the effect of the cloud-cloud collisions (\cite{Fukui15}; \cite{Saigo17}; \cite{Fukui18} submitted; \cite{Tokuda18} submitted).  The magnetohydrodynamical numerical simulations by \citet{Inoue13} showed that compression excites turbulence and amplifies field strength, leading to high-mass star formation due to the enhancement of the Jeans mass by the collision (see also \cite{Inoue18}).\\
\ \ Recent Galactic observations have indicated that high-mass stars may form via cloud-cloud collisions between intermediate mass molecular clouds.  One of the best examples is the Trifid Nebula, M20.  The total molecular mass of this system is only$\;$$\sim$$\;$10$^3$$\;$$M_{\odot}$, but it contains a single O star along with hundreds of low-mass stars. With a spatial resolution of 1$\;$pc, \authorcite{Torii11} (\yearcite{Torii11}, \yearcite{Torii17}) identified two molecular gas components with different radial velocities toward M20 with NANTEN2, Mopra and ASTE, and proposed that a recent collision between two clouds increased the effective Jeans mass and led to the formation of the O star.\\
\ \ In the present sample, we detected relatively complex velocity field toward a few sources.  Figure \ref{fig:channel_map} shows the CO intensity maps integrated with different velocity ranges.  For Target 5, there are two filamentary components having different velocities separated by $\sim$$\;$3$\;$km$\;$s$^{-1}$, and the high-mass YSO is formed at the position where the two filamentary clouds are merged.  A similar configuration is also seen toward Target 6.  At least for these clouds, there is a possibility that the interaction of molecular clouds enhanced the star formation efficiency there, and high-mass YSOs are formed therein.\\
\ \ For the other targets, multiple velocity components or significant velocity gradients are not observed, and there is no evidence of the molecular gas interaction for these sources.  However, the high-mass YSOs are formed in isolated compact clouds.  Then, what is the mechanism to initiate the high-mass star formation in such low-mass CO clouds?  Figure \ref{fig:350um} shows the {\it Herschel} 350$\;$$\mu$m images toward all the sources \citep{Meixner13}.  It is clear that the associated 350$\;$$\mu$m emission is much more extended than the CO distribution, and the high-mass YSOs tend to be located at the edge of the 350$\;$$\mu$m clouds.  Most parts of the 350$\;$$\mu$m cloud are not seen in CO, which indicates that the area is of low-density, possibly H$\;${\sc i} gas.  In this case, H$\;${\sc i} observations would be quite important to investigate the dynamical information of the surrounding gas around the present targets.  The spatial resolution of the current H$\;${\sc i} observations \citep{Kim03} is not high enough to spatially resolve the structures associated with the compact CO clouds.  The association with the edge of the 350$\;$$\mu$m clouds may indicate the existence of an effect from external sources.  Figure \ref{fig:nanten_co} shows that many of the targets are seem to be associated with H$\;${\sc i} supershells \citep{Dawson2013}.  These facts imply that the past explosive events are affecting star formation around the targets, and the future high-resolution H$\;${\sc i} observations with an angular resolution of $\lesssim$ 10 arcsec, such as ASKAP \citep{Duffy12} are highly anticipated to investigate the gas dynamics.\\

\section{Summary} \label{sec:summary}
We have carried out ALMA observations toward ten high-mass YSO candidates in isolated environments in the LMC at the highest angular resolution of 1.4 arcsecs, corresponding to a spatial resolution of 0.35$\;$pc at the distance of the LMC.  The targets are at least 200$\;$pc away from any of the NANTEN clouds whose detection limit is a few $\times$$\;$10$^4$$\;$$M_{\odot}$, and are detected by the pointed Mopra observations with a spatial resolution of 7$\;$pc.  The aim of our observations was to investigate high-mass star formation in a simple environment away from any molecular complexes, avoiding external energetic events and contamination along the same line of sight.  We have obtained the following results:

\begin{enumerate}
\item
Compact molecular clouds are associated with all the targets, indicating that these targets are bona fide YSOs formed in the compact clouds.  The masses of the clouds range from 2.8$\;$$\times$$\;$10$^2$ to 6.5$\;$$\times$$\;$10$^3$$\;$$M_{\odot}$ and the radii from 0.5 to 2.0$\;$pc.  They seem to be gravitationally bound.

\item
The excitation analyses using $^{13}$CO ($J$ = 1--0), $^{12}$CO ($J$ = 2--1) and $^{13}$CO ($J$ = 2--1) show that the density is $\sim$$\;$6$\;$$\times$$\;$10$^3$$\;$cm$^{-3}$ and the temperature is $\sim$$\;$20$\;$K.  The high density and temperature compared with high-mass star-forming clouds in the Galaxy may indicate that the photodissociation of CO is more severe due to the LMC's lower metallicity.  There is no apparent enhancement of the temperature toward the position of the YSOs, indicating that the YSOs are very young and the effect of the YSO on the gas is not significant.

\item
The star formation efficiency is calculated by assuming a standard IMF to derive the total stellar mass and is estimated to range from several percent to as high as $\sim$$\;$40\%, which means that the star formation is efficiently on-going in at least some of the observed compact molecular clumps.  

\item
Toward two targets, there is a possibility that interaction of molecular clouds enhanced the star formation activity as suggested by entangled filaments with different velocities.  While the other sources have no clear indication of the interaction of the CO gas, the interaction of lower density gas may be a cause of the high star formation activity as shown by the fact that most of the high-mass YSOs are located at the edge of the 350$\;$$\mu$m clouds.
\end{enumerate}
\clearpage 


\begin{ack}
This paper makes use of the following ALMA data: ADS/ JAO.ALMA\#2013.1.00287.S. ALMA is a partnership of the ESO, NSF, NINS, NRC, NSC, and ASIAA.  The Joint ALMA Observatory is operated by the ESO, AUI/NRAO, and NAOJ.  This work was supported by NAOJ ALMA Scientific Research Grant Numbers 2016-03B and JSPS KAKENHI (Grant No. 22244014, 23403001, 26247026, and 18K13582).  The Mopra Telescope is part of the Australia Telescope, which is funded by the Commonwealth of Australia for operation as a National Facility managed by CSIRO.  Cerro Tololo Inter-American Observatory (CTIO) is operated by the Association of Universities for Research in Astronomy Inc. (AURA), under a cooperative agreement with the National Science Foundation (NSF) as part of the National Optical Astronomy Observatories (NOAO).  M. Meixner and O. Nayak were supported by NSF grant AST-1312902.
\end{ack}

\end{document}